\documentclass[notitlepage,twocolumn,prd,nofootinbib,floatfix,preprintnumbers]{revtex4-1}
\usepackage{amsmath}         % \
\usepackage{amssymb}         %  | AMS Packages for math
\usepackage{amsfonts}        % /
\usepackage{graphicx}        % Graphics

\usepackage[left]{lineno}

\usepackage{lipsum}        % block of text (formatting test)
\usepackage{url}            % block of text (formatting test)
\usepackage{color}         % \color{...}, colored text
\usepackage{xcolor}
\usepackage{framed}        % boxed remarks
\usepackage{bbm}           % \mathbbm{1} incompatible with XeLaTeX 

\graphicspath{{figures/}}	        % set directory for figures

\renewcommand{\vec}[1]{\mathbf{#1}} % vectors are boldface

\let\oldenumerate\enumerate
\renewcommand{\enumerate}{
  \oldenumerate
  \setlength{\itemsep}{1pt}
  \setlength{\parskip}{2pt}
  \setlength{\parsep}{0pt}
}

\let\olditemize\itemize
\renewcommand{\itemize}{
  \olditemize
  \setlength{\itemsep}{1pt}
  \setlength{\parskip}{0pt}
  \setlength{\parsep}{0pt}
}

\newcommand{\be}{\begin{equation}}
\newcommand{\ee}{\end{equation}}

\newcommand{\gray}{$\gamma$-ray}
\newcommand{\grays}{$\gamma$-rays}
\newcommand{\fermilat}{{\it Fermi}--LAT}

\newcommand{\igroi}{$15^\circ \times 15^\circ$}
\newcommand{\jfac}{{\it J}-factor}

\begin{document}

\title{Dark Matter Interpretation of the \bf{{\textit{Fermi}}--LAT} Observation Toward the Galactic Center}

\author{Christopher Karwin}
\email[]{\texttt{ckarwin@uci.edu}}
\affiliation{Department of Physics and Astronomy, University of California, Irvine, CA 92697, USA}

\author{Simona Murgia}
\email[]{\texttt{smurgia@uci.edu}}
\affiliation{Department of Physics and Astronomy, University of California, Irvine, CA 92697, USA}

\author{Troy A.~Porter}
\email[]{\texttt{tporter@stanford.edu}}
\affiliation{Hansen Experimental Physics Laboratory and Kavli
Institute for Particle Astrophysics and Cosmology, Stanford University, Stanford, CA 94035, USA}

\author{Tim M.P.~Tait}
\email[]{\texttt{ttait@uci.edu}}
\affiliation{Department of Physics and Astronomy, University of California, Irvine, CA 92697, USA}

\author{Philip Tanedo}
\email[]{\texttt{flip.tanedo@ucr.edu}}
\affiliation{Department of Physics and Astronomy, University of California, Irvine, CA 92697, USA}
\affiliation{Department of Physics and Astronomy, University of California, Riverside, California 92521, USA}

\preprint{UCI-HEP-TR-2016-22}

\date{\today}

\begin{abstract}
The center of the Milky Way is predicted to be the brightest region of
\gray{s} generated by self-annihilating dark matter particles.
Excess emission about the Galactic center above predictions made
for standard astrophysical processes has been observed in \gray{} data
collected by the {\it Fermi} Large Area Telescope. It is well described
by the square of an NFW dark matter density distribution. Although other
interpretations for the excess are plausible, the possibility that it
arises from annihilating dark matter is valid. In this paper, we characterize the
excess emission as annihilating dark matter in the framework of an effective
field theory. We consider the possibility that the annihilation process
is mediated by either pseudo-scalar or vector interactions and constrain
the coupling strength of these interactions by fitting to the  {\it Fermi} Large Area Telescope
data for energies 1--100~GeV in the $15^\circ \times 15^\circ$ region
about the Galactic center using self-consistently derived interstellar emission
models and point source lists for the region.
The excess persists and its spectral characteristics favor a dark matter particle
with a mass in the range approximately from 
 50 to 190 (10 to 90) GeV 
and annihilation cross section approximately from 1$\times$10$^{-26}$ to 4$\times$10$^{-25}$ 
(6$\times$10$^{-27}$ to 2$\times$10$^ {-25}$) cm$^3$/s for pseudo-scalar (vector) interactions.
We map these intervals into the corresponding WIMP-neutron scattering cross
sections and find that the allowed range lies well below current and
projected direct detection constraints for pseudo-scalar interactions,
but are typically ruled out for vector interactions.
\end{abstract}

\maketitle

%%%%%%%%%%%%%%%%%%%%%%%%%%%%%%%%%%%%%%%%%%%%% 
\section{Introduction}
\label{sec:intro}
%%%%%%%%%%%%%%%%%%%%%%%%%%%%%%%%%%%%%%%%%%%%%

Despite the overwhelming evidence from astrophysics and cosmology that roughly 80$\%$ of the matter in our Universe is 
in the form of dark, non-baryonic particles, how this so-called dark matter (DM) fits with the
Standard Model (SM) of particle physics is currently unknown.  Determining the nature of DM is
one of the most pressing questions in the physical sciences, and a wide array of experiments are underway which
hope to shed light on its identity by observing its interactions with the better understood particles of the SM.

Indirect detection is one of the promising avenues to elucidate the nature of DM.  This method
attempts to detect and discriminate the SM particles produced by DM particle annihilations  (or decays) from those produced by conventional astrophysical processes.
\grays\ of $\sim$~GeV energies
are a particularly effective messenger because they propagate  unhindered on galactic scales, and thus
can be effectively traced back along the direction of their origin.  
In recent years, the {\em Fermi} Large Area Telescope ({\em Fermi}-LAT) has mapped out the \gray\ sky with
the highest sensitivity of space-borne detectors to date, leading to the current best limits on the annihilation cross section for
$\sim 100$~GeV DM annihilations that result in \grays.

Numerical simulations of galaxy formation offer clues as to where DM annihilation is expected to shine the most brightly.
The simulations typically predict a large concentration of DM close to the Galactic center (GC),
which smoothly falls off
with Galactocentric radius.  They also predict localized over-densities of DM, some of which correspond 
to dwarf spheroidal satellite galaxies.  Both targets provide complementary regions of interest
for DM searches.  The DM related emission from the dwarf galaxies is expected to be of lower intensity, but to be relatively free of standard astrophysical
backgrounds.
Searches for \gray\ emission from dwarf satellites of 
the Milky Way have so far shown no convincing signal of DM annihilation~\cite{Ackermann:2015zua,GeringerSameth:2011iw}.
In contrast, the GC is expected to produce a higher intensity
annihilation signal.  However, the region about the GC is strongly confused 
because of the intense interstellar emission and numerous discrete
sources of \grays\ that are summed along and through the
line-of-sight toward the GC.
The estimation of these fore-/background contributions pose a significant challenge for detection of
DM annihilation at the GC.

There seems to be an excess of \grays\ from the direction of the GC, above the expectations
from astrophysics.  This feature was first observed by Goodenough and Hooper~\cite{Goodenough:2009gk,Hooper:2010mq}, and its general features, a spatial 
morphology  remarkably consistent with predictions for a DM annihilation signal and a spectrum that peaks at a few GeV, persist  in more recent 
analyses~\cite{Hooper:2011ti,Abazajian:2012pn,Hooper:2013rwa,Gordon:2013vta,Huang:2013pda,Daylan:2014rsa,Abazajian:2014fta,Zhou:2014lva,Calore:2014xka,Abazajian:2014hsa,Calore:2014nla,Carlson:2016iis}.
The {\em Fermi}-LAT collaboration has released its own analysis~\cite{TheFermi-LAT}
of the \grays\ from the direction of the inner galaxy based on specialized  
interstellar emission models (IEMs) for estimating the fore-/background emissions, and enabling the analysis to make the
first separation of the \gray\ emission of the $\sim 1$~kpc region about the GC from the rest of the Galaxy.  Even with these IEMs, which represent
the most sophisticated modeling to date, the excess persists.
However, its spectral properties are strongly dependent on the assumed IEM, making it challenging to conclusively identify its origin.
As a result, it remains unclear whether this signal arises from DM annihilation
rather than from a currently unknown contribution from astrophysics such as
a large population of milli-second pulsars, cosmic-ray (CR) proton or electron outbursts, additional cosmic ray sources,
and/or emission from a stellar over-density in the Galactic bulge
\cite{Hooper:2013nhl,Abazajian:2014fta,Carlson:2014cwa,Petrovic:2014uda,Cholis:2014lta,Cholis:2015dea,Carlson:2016iis,Macias:2016nev}.
An interesting development is the use of statistical tools which indicate that GeV photons from 
the direction of the inner galaxy region show significantly more clustering than would be expected from
Poisson noise from smooth components \cite{Lee:2014mza,Bartels:2015aea,Lee:2015fea,McDermott:2015ydv}.  However, it remains difficult
with the current models to disentangle whether this feature represents a property of the excess itself, or unmodelled variation in the
background components \cite{Horiuchi:2016zwu}.

While it is clearly premature to claim that the GeV excess represents a confirmed signal of DM annihilation, in
this paper we extract the properties of the excess under the assumption that it does.  We make 
simultaneous fits to the parameters of generic, realistic particle physics model of DM 
annihilation together with those defining the broad characterization of the possible fore-/backgrounds
determined using the methodology of~Ref~\cite{TheFermi-LAT}.  As a result, we can compare with the expectations for such
models from direct searches for DM and colliders, finding that the null results of those searches play a significant role in
shaping the allowed parameter space.

Our work is organized as follows.  In Section~\ref{sec:analysis}, we very briefly review the methodology of 
the {\em Fermi}-LAT analysis~\cite{TheFermi-LAT}
to formulate realistic IEMs, which crucially define the fore- and backgrounds as well as
the astrophysical contributions from the GC itself.  This is followed in Section~\ref{sec:morphology} by a revisitation of some
of the most important morphological and spectral features of the signal: its centroid and whether there is evidence for two separate
components with distinct morphologies and spectra.  In Section~\ref{sec:dm}, we define
realistic flexible DM models described by effective field theories (EFTs), and
perform a maximum likelihood (ML) fit to determine the ranges of their parameters capable of describing the excess
together with the IEM parameters.
We compare the ML regions of those models to direct and collider searches for DM in Section~\ref{sec:compare}.
Section~\ref{sec:conclusions} contains our conclusions and outlook.

%%%%%%%%%%%%%%%%%%%%%%%%%%%%%%%%%%%%%%%%%%%%%
\section{Interstellar Emission Model and Analysis}
\label{sec:analysis}
%%%%%%%%%%%%%%%%%%%%%%%%%%%%%%%%%%%%%%%%%%%%%

\subsection{Data}

The analysis presented in this paper employs the
same data as used by Ref~\cite{TheFermi-LAT}: front converting events corresponding to the P7REP\_CLEAN\_V15 selection ~\cite{Ackermann:2012kna}, in the energy range 1-100 GeV, and with zenith angles less than 100$^\circ$. Exposure maps and the PSF for the pointing history of the observations were produced using the \fermilat\ ScienceTools package  (version 09-34-02)\footnote{Available at \url{http://fermi.gsfc.nasa.gov/ssc/data/analysis}}. 
Events are selected from approximately 62 months of data, from 2008-08-11 until 2013-10-15.
%We refer the interested reader to Ref~\cite{TheFermi-LAT} for a more detailed description of this dataset.
We note that for high statistics analyses such as the one presented here a notable difference  is not expected
in the results obtained with the P7REP\_CLEAN\_V15 data processing and those
processed using Pass 8 \cite{Atwood:2013rka};
this is confirmed by several previous analyses \cite{Lee:2015fea,Carlson:2016iis,Wenigerprivate}.

\subsection{Interstellar Emission Models}
\label{sec:astromodels}

The interstellar emission is the largest contribution to the
\gray\ emission toward and through the line-of-sight
 toward the GC. To separate the contribution by the Galaxy between our location and the inner 1~kpc region about the GC, 
 and that on the other side of the GC,
specialized IEMs (four in total) were developed for the Ref~\cite{TheFermi-LAT} analysis. The methodology employed templates calculated using the
well-known GALPROP CR propagation modeling code\footnote{A  description of the GALPROP code is available at \url{http://galprop.stanford.edu}} that were scaled to the
data outside of the inner \igroi\ region about the GC. Under the
assumption of Galactocentric azimuthal symmetry, these IEMs were
used to estimate the fore-/background emission over the \igroi\
region, enabling the separation. Employing this prescriptive methodology
ensures that minimal biases are introduced when fitting to the inner region. In addition, point source lists were
developed for each IEM with the properties of the individual point
sources obtained in a combined ML fit over the \igroi\ region. The construction of each IEM and its associated point-source
list/model is a critical improvement over earlier works because the
residual emission is strongly dependent on modeling both the over the region self-consistently.

The four distinct IEMs from Ref~\cite{TheFermi-LAT} are labeled:
\begin{itemize} 
\item {\it Pulsars, intensity-scaled} 
\item {\it Pulsars, index-scaled}
\item {\it OB stars, intensity-scaled}
\item {\it OB stars, index-scaled} 
\end{itemize}
The IEMs differ in the assumed distribution of the sources of CRs as tracing either the distributions
of pulsars or OB stars; 
and in the procedure employed to scale the~\gray~intensity of the fore-/background components
outside of the ~\igroi~region to the data,
either by scaling the  normalization of the model
templates for {\it intensity-scaled} IEMs, or scaling the normalization and spectral index (the latter only for gas-related templates interior to the solar circle) for the {\it index-scaled} IEMs.
Notably, it was found that the data are compatible with 
a contribution from \grays~from DM annihilation, and that the agreement between the data and the model 
significantly improves for all four IEMs when an additional component with a DM annihilation morphology
is included in the fit.  

\subsection{Analysis Procedure}

We employ the procedure developed by the \fermilat\ Collaboration in~\cite{TheFermi-LAT},
which performs a ML fit of a model consisting of
one of the four IEMs and its corresponding list of point sources to the data in the \igroi~region.  For each model, we include a 
DM annihilation contribution (described below) and perform the fit using the {\it gtlike}
package of the \fermilat\ ScienceTools.  The results of the fit are the coefficients of 
the interstellar emission components from within the the innermost $\sim$1~kpc, as well as 
those describing the DM model under consideration.
All point sources with a test statistic (defined as in~\cite{Mattox:1996zz}) $TS > 9$
are included in the model.  Their fluxes and spectra are determined by iterative fits,
with each iteration freeing the spectral parameters for a subset of point sources
in order of decreasing  {\it TS}.

%%%%%%%%%%%%%%%%%%%%%%%%%%%%%%%%%%%%%%%%%%%%%
\section{Morphology and Spectral Characteristics}
\label{sec:morphology}
%%%%%%%%%%%%%%%%%%%%%%%%%%%%%%%%%%%%%%%%%%%%%

The DM spatial distribution used in this paper is described in this section. Because \cite{TheFermi-LAT} tested spatial templates fixed at the position of Sgr A* we investigate the possibility of an offset from this location by refitting the DM spatial distribution and scanning the ML grid about the GC. 
If a large offset is found, it might challenge a DM interpretation of the excess. 
For some IEMs the DM spectrum obtained by \cite{TheFermi-LAT} extended beyond 10 GeV, but a dedicated study of the spatial distribution $>10$~GeV was not made; this is also investigated in this section.

\subsection{Dark Matter Component}
\label{sec:DM}

The results of numerical simulations for galaxy formation can broadly be described by the
Navarro, Frenk, and White (NFW) profile~\cite{Navarro:1996gj}:
\begin{equation} \label{eq1}
\rho (r) = \rho_0 \left ( \frac{r}{R_s}\right)^{-\gamma}\left(1 + \frac{r}{R_s}\right)^{\gamma - 3}
\end{equation}
 For this analysis, we use  a scale radius $R_s$ = 20 kpc, and $\rho_0$ corresponding to a local DM density $\rho_\odot$ = 0.4 $\mathrm{GeV/cm^3}$.
Two values for the inner slope $\gamma$ of the DM distribution are considered, $\gamma$ = 1, 1.2. The more cuspy distribution  $\gamma$ = 1.2  
is motivated by the possibility of halo contraction due to the influence of baryons, which are typically not
included in the simulations \cite{Diemand:2005wv}. 
The square of the NFW distribution is used as a template for DM annihilation, and we refer to it as the ``NFW profile'' (for $\gamma=1$)
or ``NFW-c'' (for $\gamma$ = 1.2). 

\subsection{NFW Centroid}
\label{sec:centroid}

%%%%%%%%%%%%%%%%%%%%%%%%%%%%%%%%%%%%%%%%%%%%%%
\begin{figure*}[th!]
\centerline{\includegraphics[width=1.0\textwidth]{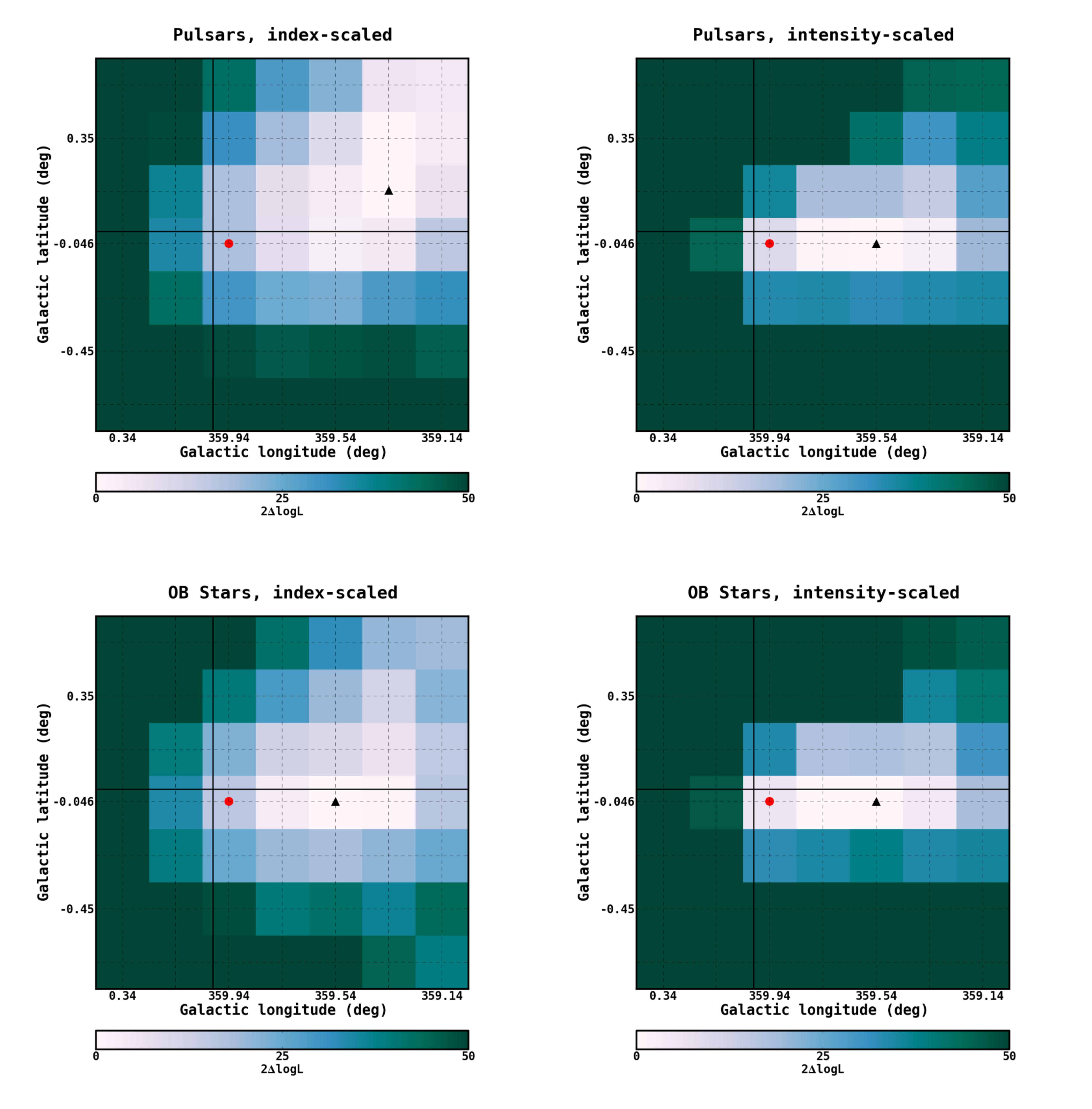}}
\caption{2$\Delta$log Likelihood as a function of the centroid position of the NFW template, as described in the text. The results are shown 
for each of the four considered IEMs, as indicated.  The  triangle and the circle indicate the position of the ML and of Sag A*, respectively.}
\label{fig:centroid}
\end{figure*}
%%%%%%%%%%%%%%%%%%%%%%%%%%%%%%%%%%%%%%%%%%%%%%%

The centroid of the Milky Way DM halo is conventionally centered at the location of 
Sgr A*. %, corresponding to the dynamical center of the Milky Way. 
Because a large offset from this location might disfavor a DM interpretation, we verify that the centroid of
the excess is sufficiently close.
An offset between the centroid of the DM halo and Sgr A* as large as approximately 2$^{\circ}$
is consistent with numerical DM simulations, with the largest offsets tending to correlate with
flatter central profiles~\cite{Kuhlen:2012qw,Lena:2014zpa}. 
An offset in the centroid position was previously 
reported in~\cite{Calore:2014xka,Linden:2016rcf}, while other studies of the GC excess have found it to be consistent with Sgr A*.

We investigate the centroid position of the excess by scanning the ML for different locations 
near Sgr A*, for each of the four IEMs.  
A power-law with exponential cut-off is employed for the spectral model, following~\cite{TheFermi-LAT}.  
The scan is performed by making the ML fit following Sec.~~\ref{sec:analysis} with the DM template centered at each point of a grid
with spacing $0.2^\circ$ centered on Sag A*.
The results of the scan are shown in Fig.~\ref{fig:centroid}, where the color scale shows the 
$\mathrm{2 \Delta log L}$ as a function of Galactic latitude and longitude. The intersections of the dotted  grid lines correspond to the points 
where the likelihood is evaluated. The  circle indicates the position of Sgr A*, and the triangle is the most likely position of the centroid
for that IEM.  We find that the centroid position is offset from Sgr A*  for all four IEMs, with the {\em Pulsars, index-scaled} 
model displaying the largest offset, both in longitude (0.6$^{\circ}$) and latitude  (0.2$^{\circ}$).
The other three models prefer an offset only in longitude (within 0.4$^{\circ}$ up to the grid accuracy).
Based on the scan, Sgr A* is not favored as the location of the NFW centroid for all four IEMs, however its position is roughly consistent with a DM interpretation for the GC excess and imperfections in the IEMs could plausibly introduce an offset.  We therefore assume for the remainder of this paper that the DM distribution is centered at Sgr~A*.

\subsection{Multiple Component Fit}
\label{sec:mcfit}

%%%%%%%%%%%%%%%%%%%%%%%%%%%%%%%%%%%%%%%%%%%%%%%%

\begin{table}[h]
\centering
\caption{Results for the multiple component fit for the {\em Pulsars, intensity-scaled} IEM.}
\label{tbl:mcfp}
\begin{tabular}{|c|c|c|}
\hline
~~~Fit components (1+2)~~~ & ~~~log L~~~ & ~~~2$\Delta$log L~~~  \\ \hline\hline
NFW + NFW  		        & -82870 & 844   \\
NFW + Gas template        & -82942 & 700  \\
NFW + 1$^\circ$ Gauss    & -82968 & 648   \\
NFW + 2$^\circ$ Gauss    & -82932 & 720  \\
NFW + 5$^\circ$ Gauss    & -82951 & 682   \\
NFW + 10$^\circ$ Gauss  & -82950  & 684   \\
NFW only 				& -82990  & 604  \\
Null hypothesis 		& -83292  & --    \\  
\hline
\end{tabular}
\end{table}

\begin{table}[h]
\centering
\caption{Results for the multiple component fit for the {\em OB stars, intensity-scaled} IEM.}
\label{tbl:mcfob}
\begin{tabular}{|c|c|c|}
\hline
~~~Fit components (1+2)~~~ & ~~~log L~~~ & ~~~2$\Delta$log L~~~  \\ \hline\hline
NFW + NFW  		            & -82972 & 914   \\
NFW + Gas template       	    & -83068 & 722  \\
NFW + 1$^\circ$ Gauss        & -83096 & 666   \\
NFW + 2$^\circ$ Gauss        & -83065 & 728  \\
NFW + 5$^\circ$ Gauss        & -83147 & 564   \\
NFW + 10$^\circ$ Gauss      & -83111   & 636   \\
NFW only 				    & -83099  & 660  \\
Null hypothesis 	             & -83429 & --    \\  
\hline
\end{tabular}
\end{table}

%%%%%%%%%%%%%%%%%%%%%%%%%%%%%%%%%%%%%%%%%%%%%%%

 %%%%%%%%%%%%%%%%%%%%%%%%%%%%%%%%%%%%%%%%%%%%%%
\begin{figure*}[th!]
\centerline{\includegraphics[width=0.85\textwidth]{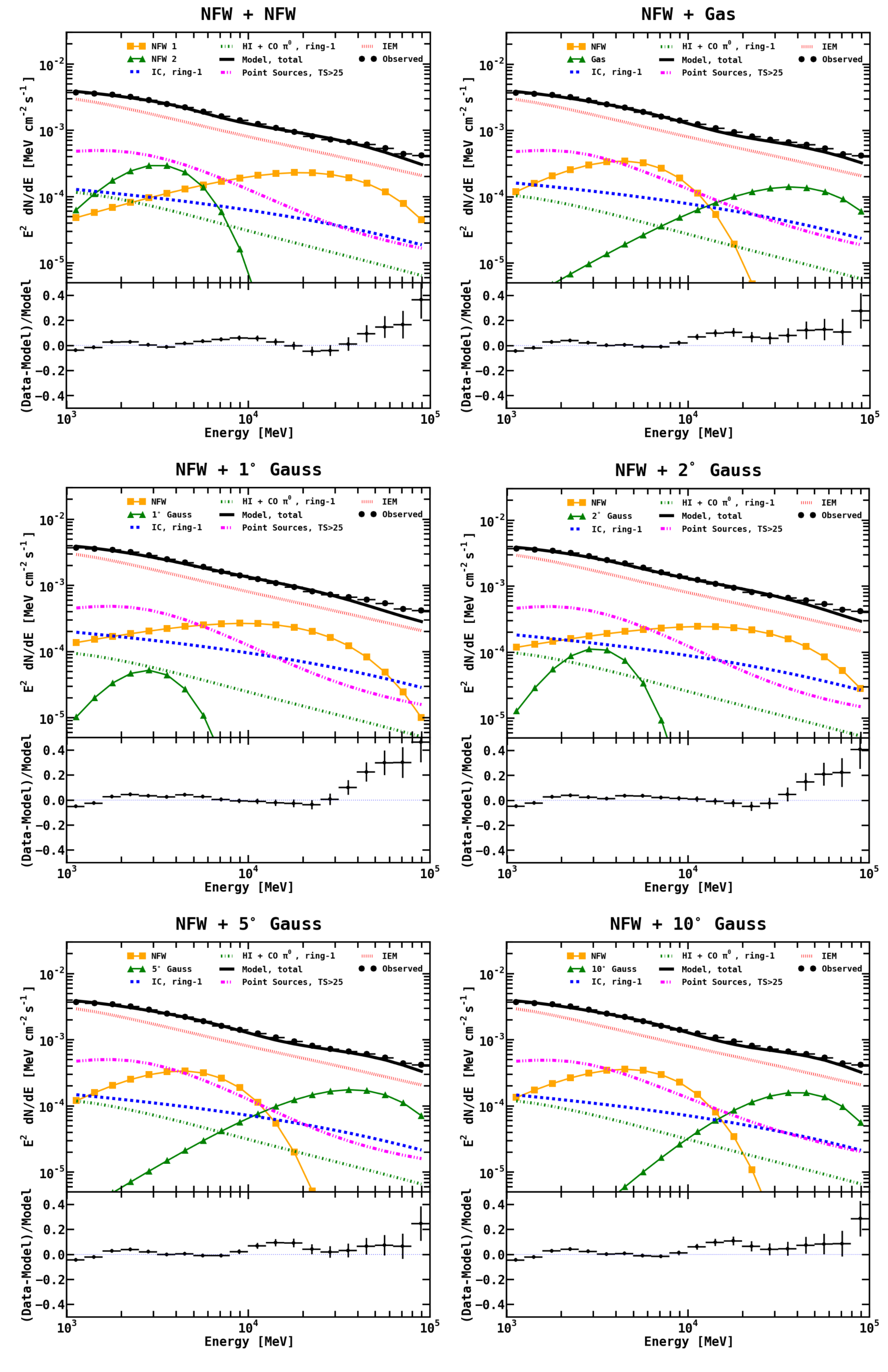}}
\caption{Differential fluxes (broken down into components, as indicated) 
integrated over the \igroi~region for the two component fits, along with their fractional residuals,
for the {\em Pulsars, intensity-scaled} IEM.
}
\label{fig:mcf}
\end{figure*}
%%%%%%%%%%%%%%%%%%%%%%%%%%%%%%%%%%%%%%%%%%%%%%

Whether the high-energy tail ($>10$~GeV) of the GeV excess spectrum is related to that at lower energies remains an open issue.
In~\cite{TheFermi-LAT},  the excess emission above 10 GeV is most prominent in the intensity-scaled IEMs. 
For the index-scaled variants however, it is largely attributed to interstellar emission (see also \cite{Daylan:2014rsa}).
The origin of the $> 10$~GeV excess has been previously investigated by several studies.
In \cite{Horiuchi:2016zwu}, the excess emission above 10 GeV is found 
to cut off in the innermost few degrees about the GC (unlike the excess at a few GeV) and therefore to have a different spatial morphology;
secondary emission from unresolved millisecond pulsars is proposed as an interpretation. 
In \cite{Linden:2016rcf}, the  excess emission above 10 GeV is found to have a similar radial profile as the peak emission.
Ref~\cite{Linden:2016rcf} also discusses the interplay with the Fermi Bubbles, although the bubble morphology close to the Galactic plane is uncertain. 

Here we investigate the morphology of the $>10$~GeV excess emission present
for the {\em Pulsars} and {\em OB stars, intensity-scaled} IEMs. 
We perform a ML fit over the 1-100~GeV energy range with two components
to model the GC excess:
an NFW template; and a second component that has either an
NFW,  gas, or a 2D gaussian (with half-width, half maximum of $1^\circ$, $2^\circ$,  $5^\circ$, or $10^\circ$) morphology. 
These are the same templates that were employed by~\cite{TheFermi-LAT}.
Six template combinations for the two intensity-scaled models are therefore tested.
The spectrum for each template is modeled as a power law with an exponential cutoff function. 
The ML fit  is performed iteratively, as described in section~\ref{sec:analysis}, and the results  are shown in 
Tables~\ref{tbl:mcfp} and~\ref{tbl:mcfob} for the {\em Pulsars} and {\em OB stars, intensity-scaled} IEM, respectively.  The NFW + NFW combination is favored over all of the others considered, for
both IEMs.

In Fig.~\ref{fig:mcf} the differential fluxes integrated over the \igroi~region for the two component fits, along with the fractional residuals, are shown
for the {\em Pulsars, intensity-scaled} model.
The contribution to the flux from each of the two spatial components and the IEM are shown, 
with the IEM broken down into the contributions from inverse Compton (IC), $\pi^0$ emission from the inner $\sim$ 1~kpc (``ring 1" in the legend), and from the point sources.
For each of the six combinations we consider, the low energy excess is better described by an NFW morphology. 
The more peaked 2D gaussian templates ($1^\circ$  and  $2^\circ$)  have spectra that peak in the few GeV energy range and cutoff at higher energies. 
Note that their contribution is  always well below the contribution assigned to the NFW template. 
On the other hand, the spectra for the broader 2D gaussian templates ($5^\circ$  and  $10^\circ$) are more prominent at higher energies, suggesting that the high-energy 
tail of the GeV excess is consistent with an extended component in the region. 
The NFW morphology, which is peaked towards the GC and broadly extended in the region, is better suited to model the 
excess emission over the full energy range compared to the other options we have considered.
However, due to the limitations of the IEMs together with the limited statistics at the higher energies, it is difficult to conclude
decisively whether or not the high-energy tail is a true feature of the GC excess.  Given the current preference for a single NFW morphology
for both low and high energy components, we include the full energy range when comparing with the DM scenarios  in Section~\ref{sec:dm} below.

%%%%%%%%%%%%%%%%%%%%%%%%%%%%%%%%%%%%%%%%%%%%%
\section{Dark Matter Interpretation}
\label{sec:dm}
%%%%%%%%%%%%%%%%%%%%%%%%%%%%%%%%%%%%%%%%%%%%%

In this section we fit the parameters of particle physics models of DM, together with the parameters describing
the fore-/backgrounds, extracting a comprehensive DM interpretation of the GC excess.  
As described in more detail below, we employ a parameterization 
of the DM particle physics model which  allows for distinct annihilation rates into up-type quarks, down-type quarks, and leptons. 
Our parametrization has more flexibility than the often-considered annihilation into a single channel of SM particles and, in this sense,  is better able to
capture a wider array of realistic particle physics models for DM annihilation than those typically used
in indirect searches.

\subsection{EFT Description of Dark Matter Interactions}
\label{sec:eft}

We consider two representative EFTs that describe the DM interactions with the SM fermions.
These theories form part of a universal set of operators to which any theory of DM flows at low energies, well below
the masses of the particles responsible for communicating between the SM and the dark 
matter \cite{Beltran:2008xg,Cao:2009uw,Beltran:2010ww,Goodman:2010ku,Goodman:2010qn,Kumar:2013iva}.
Such models have previously been considered to describe the GC excess \cite{Alves:2014yha,Liem:2016xpm}.
More generalized constructions are employed here, and their parameters are fit 
together with the IEM parameters as described in Section~\ref{sec:analysis}.
Of course, models with light mediators are also interesting, and worthy of investigation in their own right 
\cite{Boehm:2014hva,Abdullah:2014lla,Martin:2014sxa,Berlin:2014pya,Balazs:2014jla,Ko:2014loa,Carpenter:2016thc,Escudero:2016gzx}.  
We leave exploration of such theories for future work.

Both of our considered EFTs are chosen such that they mediate $s$-wave (velocity-unsuppressed) annihilation, because a $p$-wave
annihilation mechanism would require such strong interactions to overcome the innate
$v^2 \sim 10^{-4}$ suppression that it is likely to already be ruled out by direct and/or collider searches.  We further restrict them to follow the
principle of minimal flavor violation (MFV) \cite{D'Ambrosio:2002ex}, such that the most stringent constraints from flavor-violating observables
are mitigated by small Yukawa interactions.  We consider models containing either pseudo-scalar or vector Lorentz structures
described by Lagrangians ${\cal L}_{\rm ps}$ and ${\cal L}_{\rm vec}$ (respectively, in the fermion mass basis),
\begin{eqnarray}
{\cal L}_{\rm ps} & = &  \overline{\chi} \gamma_5 \chi 
\times  \label{eq:ps} \\ & & 
\sum_i \left\{ \frac{m_{u_i}}{\Lambda_u^3} ~\overline{u}_i \gamma_5 u_i
+ \frac{m_{d_i}}{\Lambda_d^3} ~\overline{d}_i \gamma_5 d_i  
+ \frac{m_{\ell_i}}{\Lambda_\ell^3} ~\overline{\ell}_i \gamma_5 \ell_i \right\}, \nonumber \\
{\cal L}_{\rm vec} & = &  \overline{\chi} \gamma^\mu \chi 
\times \label{eq:vec} \\ & & 
\sum_i \left\{ \frac{1}{\Lambda_u^2} \overline{u}_i \gamma_\mu u_i
+ \frac{1}{\Lambda_d^2} \overline{d}_i \gamma_\mu d_i  
+ \frac{1}{\Lambda_\ell^2} \overline{\ell}_i \gamma_\mu \ell_i \right\}, \nonumber
\end{eqnarray}
where $i=1,2,3$ is the sum over fermion flavor with the indicated
relative weighting of $m_{f_i}$ $(1)$ for the pseudo-scalar (vector) interaction
types, as dictated by the leading terms consistent with MFV.  The $\Lambda_{u,d,\ell}$ are parameters with dimensions
of energy which specify the separate
interaction strengths between the DM and up-type quarks, down-type quarks, and charged leptons.  
Together with the DM mass, $m_\chi$,
these coefficients specify the point in parameter space for the DM model.  They represent generalizations (in that they allow the couplings
of up-type and down-type quarks and leptons to vary independently) of the commonly considered interactions D4 and D5
used in DM searches via direct detection and at colliders~\cite{Goodman:2010ku}.

\subsection{\gray~Flux from Dark Matter Annihilation}

The interactions in both the pseudo-scalar and vector models defined in Eqs.~(\ref{eq:ps},\ref{eq:vec}) lead to cross sections for
a pair of DM particles to annihilate $\chi \overline{\chi} \rightarrow f \overline{f}$ (where $f$ is any SM fermion):
\begin{eqnarray}
\langle \sigma_f v \rangle_{\rm ps} &=& \frac{N_f m_f^2 m_\chi^2}{\Lambda_f^6 \pi} \sqrt{ 1 - \frac{m_f^2}{m_\chi^2} } + {\cal O}(v^2), 
\label{eq:sigmaav} \\
\langle \sigma_f v \rangle_{\rm vec} &=& \frac{N_f (2 m_\chi^2 + m_f^2)}{\Lambda_f^4 \pi} \sqrt{ 1 - \frac{m_f^2}{m_\chi^2} } + {\cal O}(v^2),
\label{eq:sigmavec}
\end{eqnarray}
where $\langle \cdot \rangle$ indicates averaging over the DM velocity profile,
$N_f = 3$ (1) for quarks (leptons) counts their color degrees of freedom, and $\Lambda_f$ is the appropriate $\Lambda_{u,d,\ell}$ for the
fermion under consideration.  The inclusive cross section for annihilation into up-type quarks, down-type quarks, and 
charged leptons is the sum of the
individual cross sections for all three flavors of each fermion type, and the total cross section $\langle \sigma v \rangle$ is the sum of the
three inclusive cross sections.  In presenting results, we typically trade the three parameters $\Lambda_{u,d,\ell}$ for $\langle \sigma v \rangle$
and the fractional cross sections $f_u$, $f_d$, and $f_\ell$ (with $f_u + f_d + f_\ell = 1$).  It is easy to map these back into the
$\Lambda_{u,d,\ell}$ parameters using the appropriate single channel cross section from Eqs.~(\ref{eq:sigmaav}) and (\ref{eq:sigmavec}).

%%%%%%%%%%%%%%%%%%%%%%%%%%%%%%%%%%%
\begin{figure*}
\includegraphics[width=1.0\textwidth]{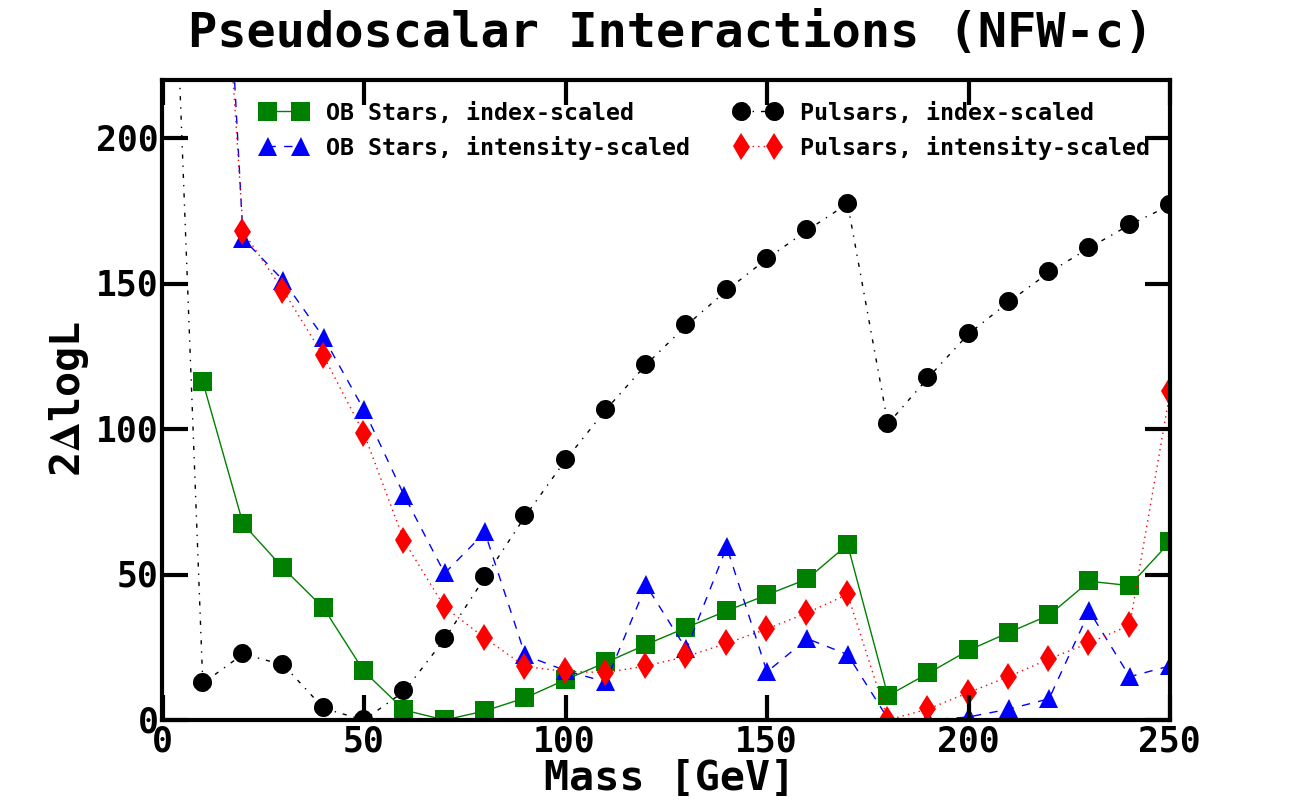}
\caption{Likelihood ($\mathrm{2 \Delta logL}$)  as a function of the DM mass  for the
pseudo-scalar interaction model with NFW-c morphology. Results are shown for all four IEMs, as indicated.}
\label{fig:pslike}
\end{figure*}
%%%%%%%%%%%%%%%%%%%%%%%%%%%%%%%%%%%%

The \gray~intensity and spectrum from DM annihilation 
is constructed by summing over all of the annihilation channels:
\begin{equation} 
\label{eq7}
\frac{dN_\gamma}{dE} = \sum_f
\frac{ \langle \sigma_f v \rangle}{4 \pi \eta ~ m_\chi ^2} \frac{dN^f_\gamma}{dE} \times \int_{\Delta\Omega} d\Omega^\prime \int_{los} ds \ \rho ^2 (r(s, \psi)),
\end{equation}
where 
${dN^f_\gamma}/dE$ is the number of $\gamma$ rays per annihilation
into the $f \overline{f}$ channel, generated from the PPPC 4 DM ID package~\cite{Cirelli:2010xx}
based on fits to Pythia 8.1 \cite{Sjostrand:2007gs}, and
$\eta$ = 2(4) for Majorana (Dirac) DM.
The integral is the \jfac, obtained by integrating the DM density $\rho^2(\vec{x})$ corresponding to either an NFW or NFW-c distribution,
Eq.~(\ref{eq1}), over the line of sight ($los$) in direction $\psi$. 

To determine the preferred DM model parameters for each IEM, we fix the DM mass in the range from 10 -- 250~GeV
in 10 GeV increments.  For each mass hypothesis the analysis procedure of Section~\ref{sec:analysis} determines the  fitted
values of the DM model parameters $f_u$, $f_d$, and $f_\ell$, along with the coefficients of 
the interstellar emission components from within the  innermost $\sim$1~kpc and point sources, as usual.  
We repeat this scan for both NFW
and NFW-c annihilation morphologies and for both the pseudo-scalar and vector models described above. We find that the DM component is 
detected with high statistical significance for all IEMs, and  for pseudo-scalar as well as vector interactions.
The likelihood values  for pseudo-scalar interactions are summarized in Table~\ref{tbl:dmfit}.

%%%%%%%%%%%%%%%%%%%%%%%%%%%%%%%%%%%%%%%%%%%%%%%%

\begin{table*}[t]
\centering
\caption{Likelihood ($\mathrm{log~L}$) values for all IEMs  for pseudo-scalar interactions and for  NFW and NFW-c templates.}
\label{tbl:dmfit}
\begin{tabular}{|c|c|c|c|}
\hline
~~~IEM~~~ & ~~log L (null hypothesis)~~ & ~~log L (NFW)~~ & ~~log L (NFW-c)~~ \\ \hline\hline
{\em Pulsars, index-scaled }        & -82926 & -82738 & -82739   \\
{\em Pulsars, intensity-scaled }   & -83292 & -82965 & -82956   \\
{\em OB stars, index-scaled }     & -82993 & -82779 & -82806   \\
{\em OB stars, intensity-scaled } & -83429 & -83081 & -83117   \\
\hline
\end{tabular}
\end{table*}

%%%%%%%%%%%%%%%%%%%%%%%%%%%%%%%%%%%%%%%%%%%%%%%%

\subsection{Results for Pseudo-scalar Interactions}
\label{sec:spin0}

In Fig.~\ref{fig:pslike}, we display the likelihood profile as a function of the DM mass for each of the IEMs 
for the NFW-c annihilation morphology.  The results for the NFW morphology are qualitatively similar.
Each of the four IEMs shows a clear preference for particular DM masses, but there is considerable variation between them,
 with the index-scaled models favoring a mass around $\sim$ 50~GeV, while the intensity-scaled models favor higher masses $\sim$ 200~GeV. 
The results are consistent with the results  
obtained by~\cite{TheFermi-LAT}, where  the spectrum of the GC excess for the index-scaled IEMs displays a lower energy cutoff compared to the intensity-scaled IEMs. 
The spectra we consider here correspond to motivated DM scenarios, in contrast with the simpler assumptions 
made for the spectral model by~\cite{TheFermi-LAT}.

%%%%%%%%%%%%%%%%%%%%%%%%%%%%%%%%%
\begin{figure*}
\includegraphics[width=1\textwidth]{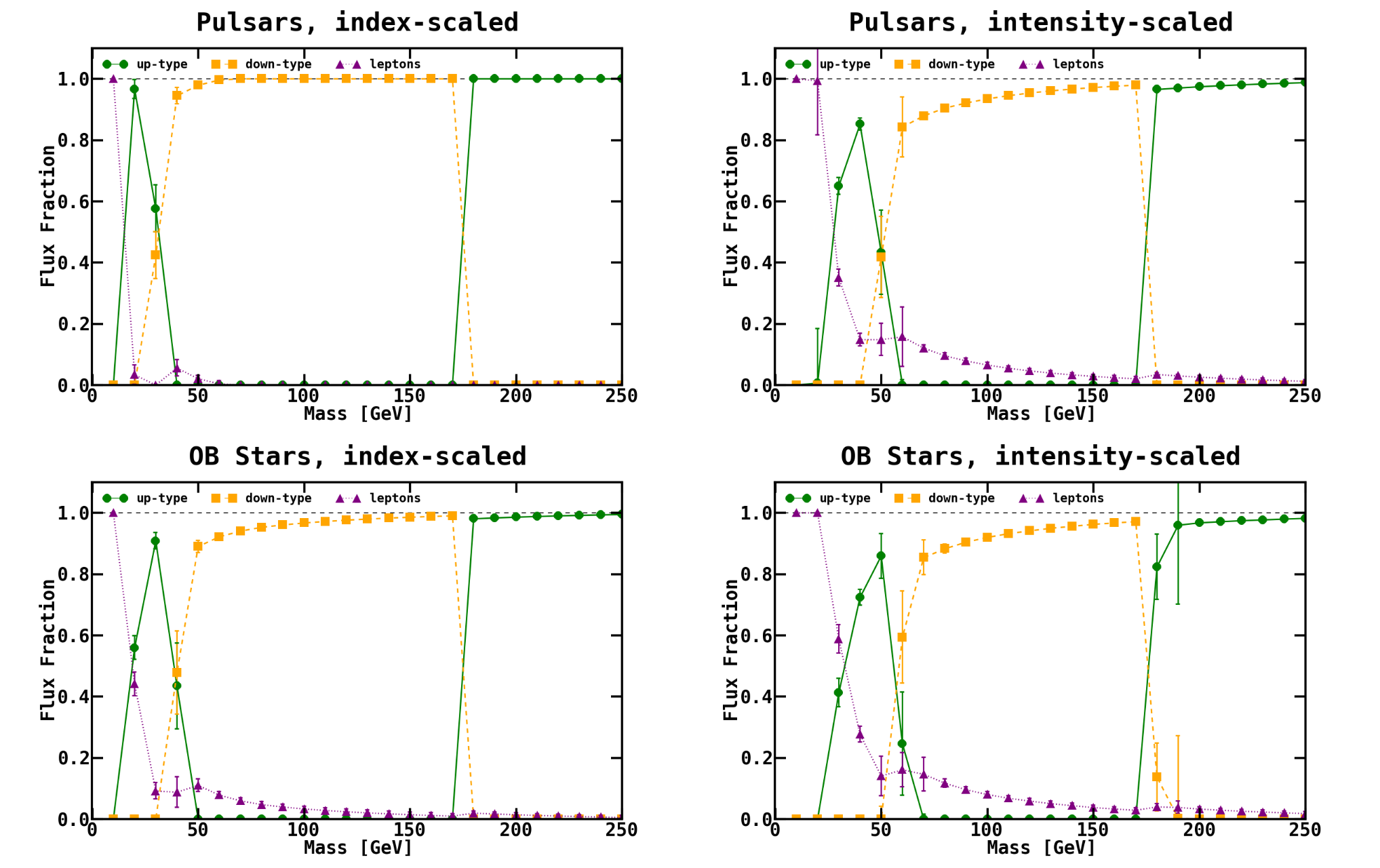}
\caption{Flux fraction for annihilation into up-type quarks, down-type quarks, and charged leptons, 
for the pseudo-scalar interaction model with NFW-c morphology. Results are shown for all four IEMs, as indicated.}
\label{fig:dmfrac}
\end{figure*}
%%%%%%%%%%%%%%%%%%%%%%%%%%%%%%%%%%%

In Fig.~\ref{fig:dmfrac}, we present the ML fractions into the three annihilation channels as a function of the DM mass, for
each of the IEMs with the NFW-c annihilation morphology.  These also vary considerably from one IEM to another, and are characterized
by one channel or another typically dominating at any given DM mass hypothesis:
charged leptons at lower masses $\sim10-20$~GeV;
down-type quarks in the range $\sim 50 -170$ GeV; and  
up-type quarks above 180~GeV and at lower masses $\sim20-40$~GeV.   
The lepton flux declines steeply above $\sim$20~GeV, and its contribution to the flux is smaller for the index-scaled models 
({\em Pulsars} in particular) compared to the intensity-scaled ones.  This reflects in part
the lower energy cutoff of the GC excess spectrum for the index-scaled models and the harder \gray\ spectra produced
by charged leptons compared to quarks.
Also of note is the sharp transition from annihilation into
down-type quarks to up-type quarks at the top mass threshold, $\sim175$~GeV.   This follows because
the pseudo-scalar model annihilations are dominated by the heaviest quark kinematically accessible, and top quarks produced
close to at rest decay into $\sim 60$~GeV bottom quarks, corresponding to the ML region at $m_\chi \sim 50$~GeV.

The best-fit DM mass for the {\em Pulsars} ({\em OB stars}) index-scaled IEM is 
$\mathrm{50_{-10}^{+10}}$ GeV ($\mathrm{70_{-10}^{+15}}$ GeV), 
and in both cases annihilation is predominantly into bottom 
quarks\footnote{The grid spacing is taken into account in the quoted uncertainties on the DM mass.}.
These results are compatible with the findings of previous studies \cite{Agrawal:2014oha,Balazs:2016dcg}
interpreting the spectrum of the excess as presented in Ref.~\cite{TheFermi-LAT}.
The intensity-scaled IEMs favor higher DM masses, $\mathrm{180_{-5}^{+15}}$ GeV and $\mathrm{190_{-15}^{+25}}$ GeV, for the 
{\em Pulsars} and {\em OB stars} variants, respectively, and primarily favor annihilation into top quarks. 
We note that the likelihood profile for the {\em OB stars, intensity-scaled} IEM is rather flat around the minimum, which yields a higher 
uncertainty in the best-fit DM mass, compared to the other IEMs. The  uncertainties on the flux fractions into up-type and down-type
quarks in this mass range are also somewhat larger.

%%%%%%%%%%%%%%%%%%%%%%%%%%%%%%%%%%%%%%
\begin{figure*}
\centerline{\includegraphics[width=1.0\linewidth]{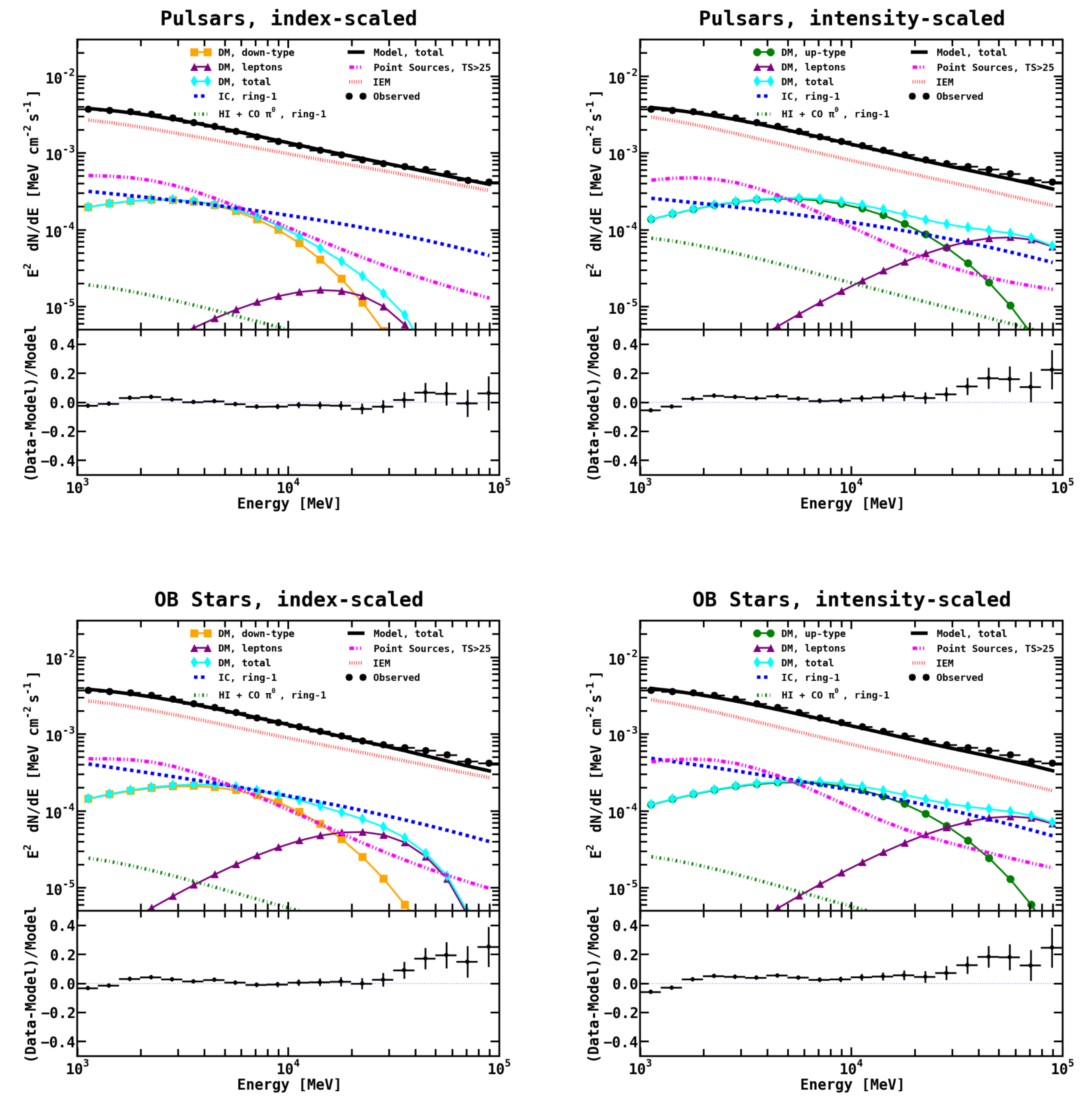}}
\caption{Differential fluxes (broken down into components, as indicated) 
integrated over the \igroi~region  and corresponding fractional residuals for pseudo-scalar interactions and for the four IEMs. 
}
\label{fig:spec}
\end{figure*}
%%%%%%%%%%%%%%%%%%%%%%%%%%%%%%%%%%%%%%

The differential fluxes for the ML model (and the data points) are shown for each IEM in Fig.~\ref{fig:spec}.  
Individual model components are displayed  separately, including the contribution to the DM flux from each annihilation final state, as well as their sum.
The contribution from each DM annihilation channel illustrates the fact that 
the integrated DM flux originates primarily from annihilations into quarks with the harder spectrum from annihilation into leptons
becoming important at higher energies, particularly for the intensity-scaled IEMs.
The \gray~emission correlated with gas from the innermost $\sim$1~kpc is sub-dominant
in the region.
Fig.~\ref{fig:spec} also shows the fractional residuals as a function of energy. 
The agreement between data and model is at the level of a few \% or better up to $\sim 30$~GeV for all IEMs, and is 
generally worse at higher energies for all but the {\em Pulsars, index-scaled} IEM. 
It is plausible that the energy cutoff at the DM mass in the annihilation spectrum 
limits its ability to describe the excess at the higher energies while simultaneously providing a good fit to the data in the few GeV range. 
We note that the fractional residuals based on realistic DM models including up-type, down-type, and lepton final states generally improve  (for the same number of free parameters) over the results in~\cite{TheFermi-LAT} based on a power law
with exponential cutoff spectrum.

%%%%%%%%%%%%%%%%%%%%%%%%%%%%%%%%%%%%%%%%%%%%
\begin{figure*}
\centerline{\includegraphics[width=1.0\textwidth]{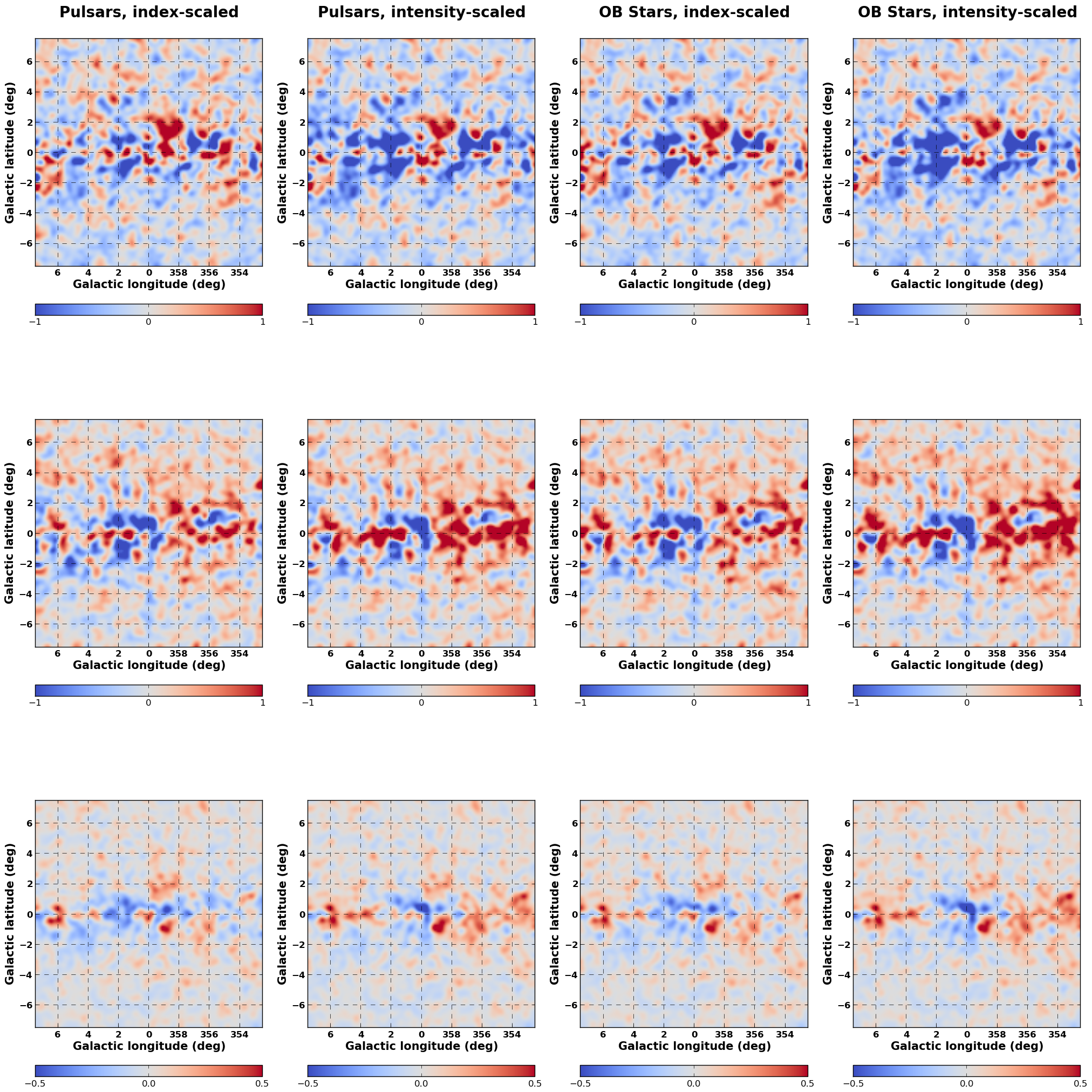}}
\caption{Residuals (data -- model) in three  energy bands, for the four IEMs. The  rows correspond to the range  1 - 1.6 GeV (top),  1.6 - 10 GeV (center), and 10 - 100 GeV (bottom). The columns, going from left to right are: {\em Pulsars, index-scaled}; {\em Pulsars, intensity-scaled}; 
{\em OB stars, index-scaled}; {\em OB stars, intensity-scaled}.}
\label{fig:res}
\end{figure*}
%%%%%%%%%%%%%%%%%%%%%%%%%%%%%%%%%%%%%%%%%%%%

Residual count (data-model) maps  are shown in Fig.~\ref{fig:res} for the energy bands $1-1.6$, $1.6-10$, and $10-100$~GeV, 
for each IEM. Structured excesses and deficits remain that may be attributed to imperfect modeling of the interstellar emission. 
Because of this, we do not rule out the DM models corresponding to IEMs with 
larger fractional residuals as these discrepancies might be explained by limitations in the IEMs. 
There is better agreement with the data when the DM spectrum is modeled with power law functions in 10 
independent energy bins as done in~\cite{TheFermi-LAT}; perhaps unsurprising given the
larger number of free parameters for the spectral model.

%%%%%%%%%%%%%%%%%%%%%%%%%%%%%%%%%%%%%%%%%%%%
\begin{figure*}
\includegraphics[trim = 90 0 0 0,scale=0.29]{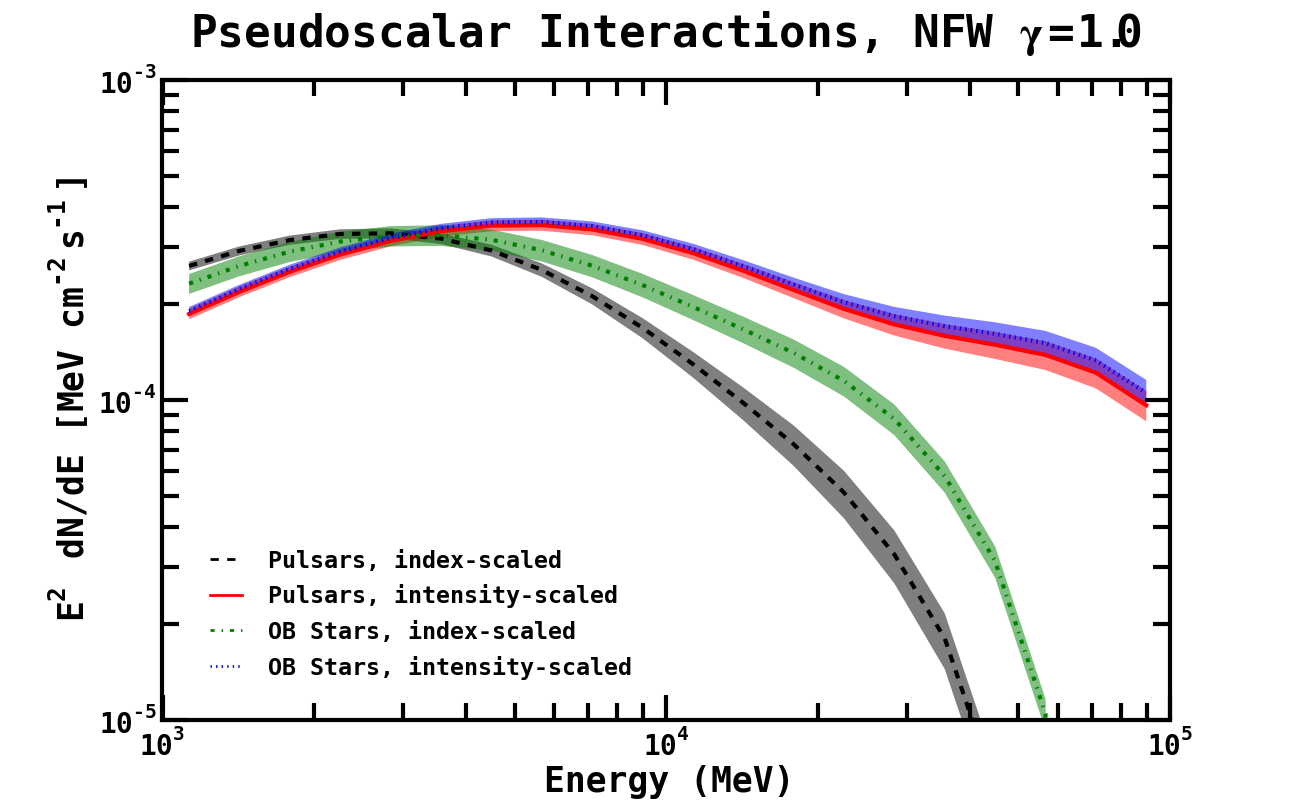} ~~~~
\includegraphics[trim = 90 0 0 0,scale=0.29]{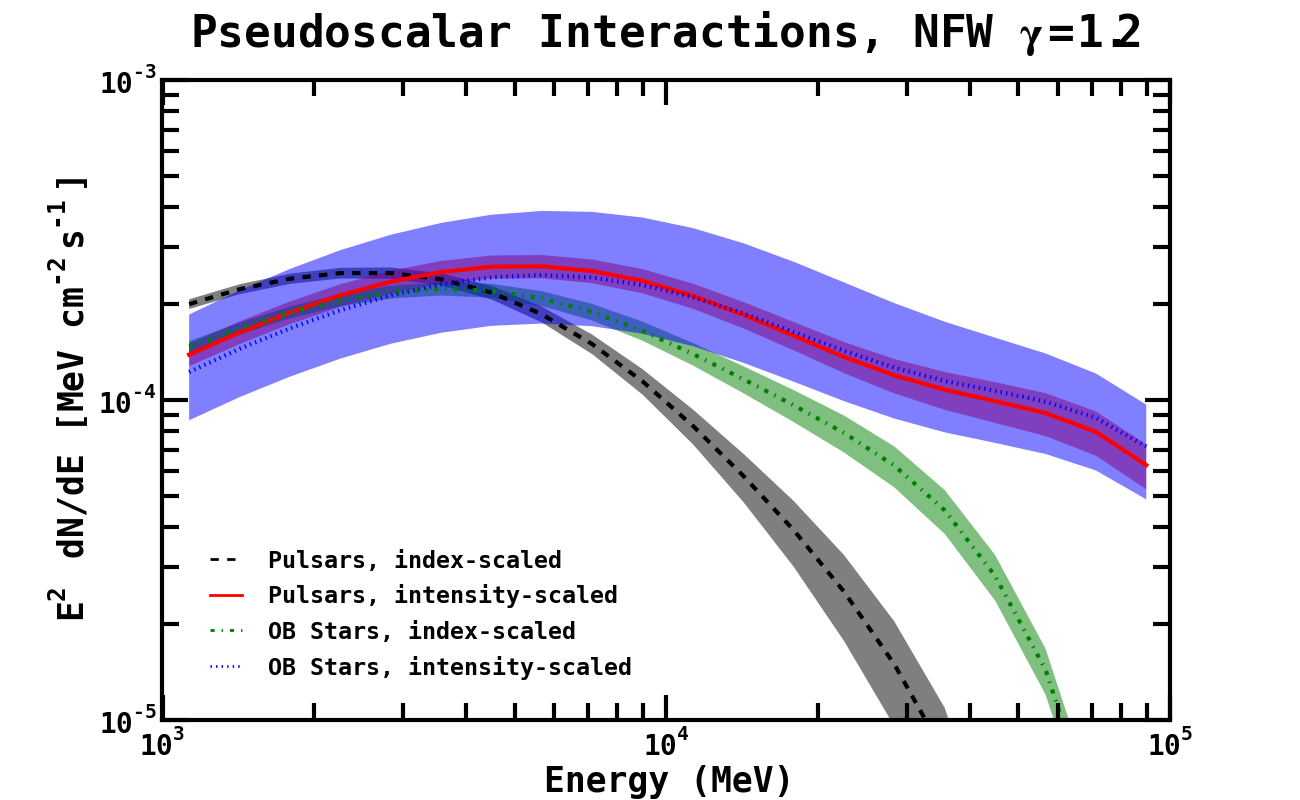}
\caption{Differential flux integrated over the \igroi~region  for the DM component for pseudo-scalar interactions,  NFW and  NFW-c profiles,  for all four IEMs, as indicated. The bands represent the fit uncertainties on the normalization.}
\label{fig:dmflux}
\end{figure*}
%%%%%%%%%%%%%%%%%%%%%%%%%%%%%%%%%%%%%%%%%%%%%

The differential flux from the total DM annihilation component for both profiles (NFW, NFW-c) and all four IEMs are summarized in Fig.~\ref{fig:dmflux}. 
The bands represent the 1$\sigma$ fit uncertainty on the flux summing the up-type, down-type, and lepton final states. 
For the index-scaled variants of the IEMs, the spectrum peaks at a few GeV, while for the intensity-scaled counterparts the peak 
shifts to higher energies. This is consistent with the requirement that the high energy tail in the spectrum for the intensity-scaled IEMs, 
predominantly from annihilations into leptons, has to cutoff at the  same energy (corresponding to the DM mass) as the contribution to the flux 
from annihilations into up-type and down-type quarks, which dominate the DM flux at lower energies.   
Finally, we note that the flux for NFW-c  profile is smaller compared to the NFW profile. 
As a consequence, a simple rescaling based on $J-$factors when comparing fit results obtained with different profiles is not accurate, 
as the flux assigned to the DM component has a dependence on the specific morphology.

%%%%%%%%%%%%%%%%%%%%%%%%%%%%%%%%%%%%%%%%%%%%%%
\begin{figure*}
\includegraphics[width=1.0\textwidth]{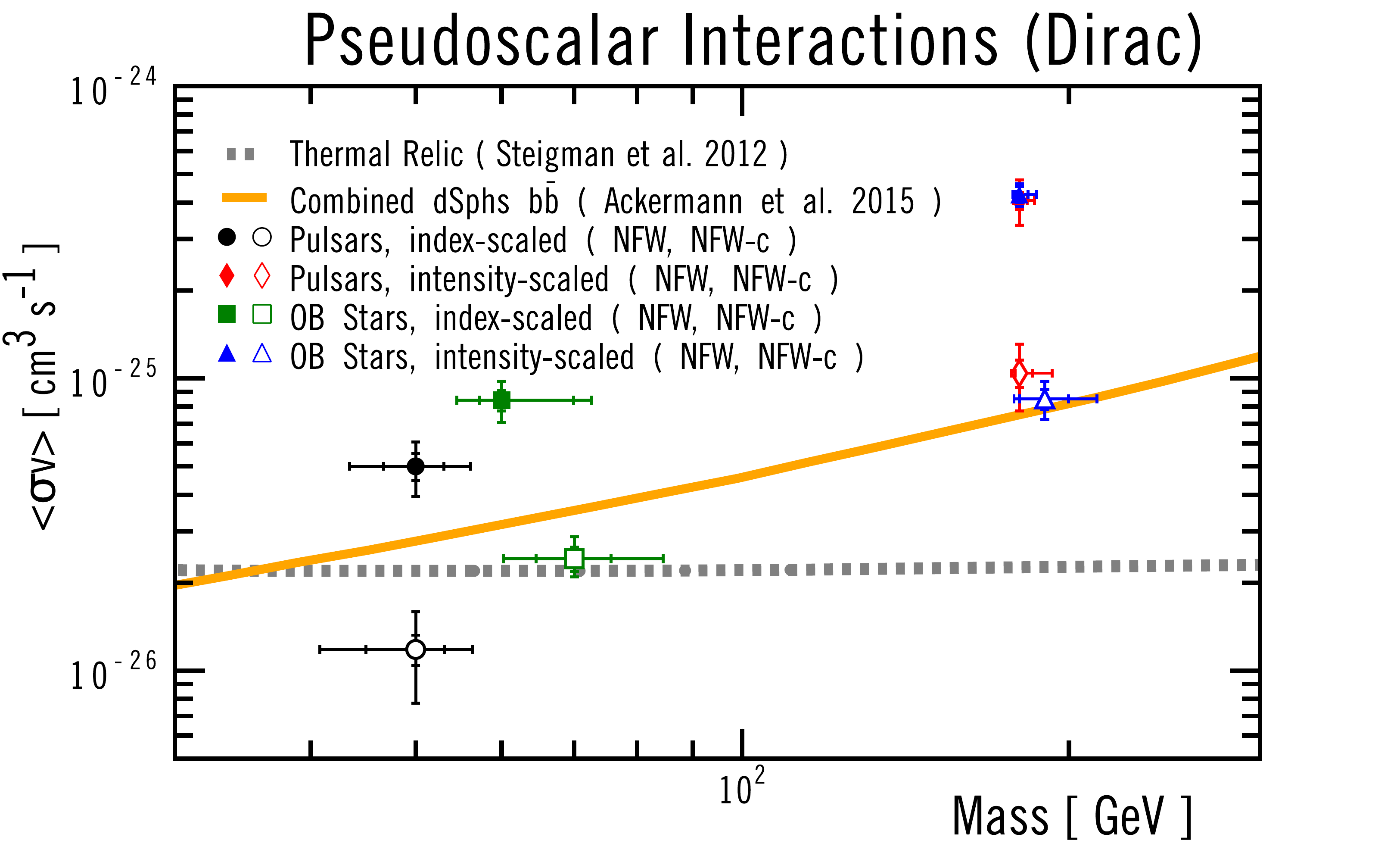}
\caption{Masses and cross sections for pseudo-scalar interaction models
(including one and two sigma uncertainties as the tick marks) for NFW and NFW-c DM profiles,
and the four IEMs, as indicated.  Also shown are the cross sections saturating the standard thermal relic density (grey dashed line)
and the \fermilat\ $95\%$ C.L. bounds from dwarf spheroidal galaxies, assuming $100\%$ annihilation into $b \overline{b}$.}
\label{fig:xsec}
\end{figure*}
%%%%%%%%%%%%%%%%%%%%%%%%%%%%%%%%%%%%%%%%%%%%%%
 
We translate the DM template flux for each IEM
into the inclusive annihilation cross section, with the
results shown in Fig.~\ref{fig:xsec}.
Also shown for comparison
is the $\langle \sigma v \rangle$ predicting saturation the measured DM relic density for a standard
cosmology \cite{Steigman:2012nb}.
The results for the index-scaled models are comparable to those found in most of the earlier studies of the GeV excess
\cite{Hooper:2010mq,Abazajian:2012pn,Hooper:2013rwa,Gordon:2013vta,Huang:2013pda,Daylan:2014rsa,Abazajian:2014fta,Zhou:2014lva,Calore:2014xka,Abazajian:2014hsa,Calore:2014nla,Carlson:2016iis}.
The intensity-scaled models however are consistent with larger DM masses and cross sections, 
as first discussed in \cite{Agrawal:2014oha}, based on the spectra from~\cite{TheFermi-LAT}.  

\subsection{Results for Vector Interactions}
\label{sec:spin1}

%%%%%%%%%%%%%%%%%%%%%%%%%%%%%%%%%%%
\begin{figure*}
\includegraphics[width=1\textwidth]{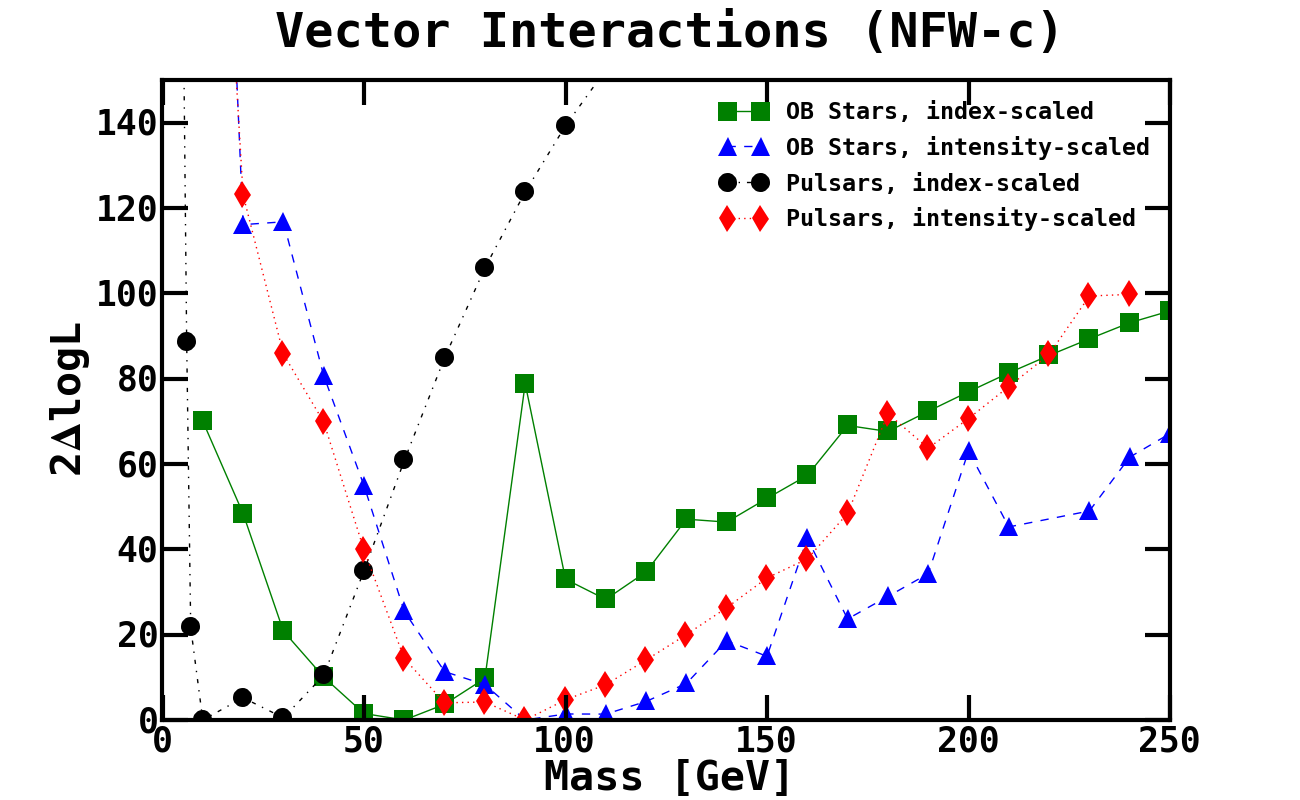}
\caption{Likelihood ($\mathrm{2 \Delta logL}$)  as a function of the DM mass  for the
vector interaction model with NFW-c morphology. Results are shown for all four IEMs, as indicated.}
\label{fig:pslikespin1}
\end{figure*}
 %%%%%%%%%%%%%%%%%%%%%%%%%%%%%%%%%

 %%%%%%%%%%%%%%%%%%%%%%%%%%%%%%%%%
\begin{figure*}
\includegraphics[width=1\textwidth]{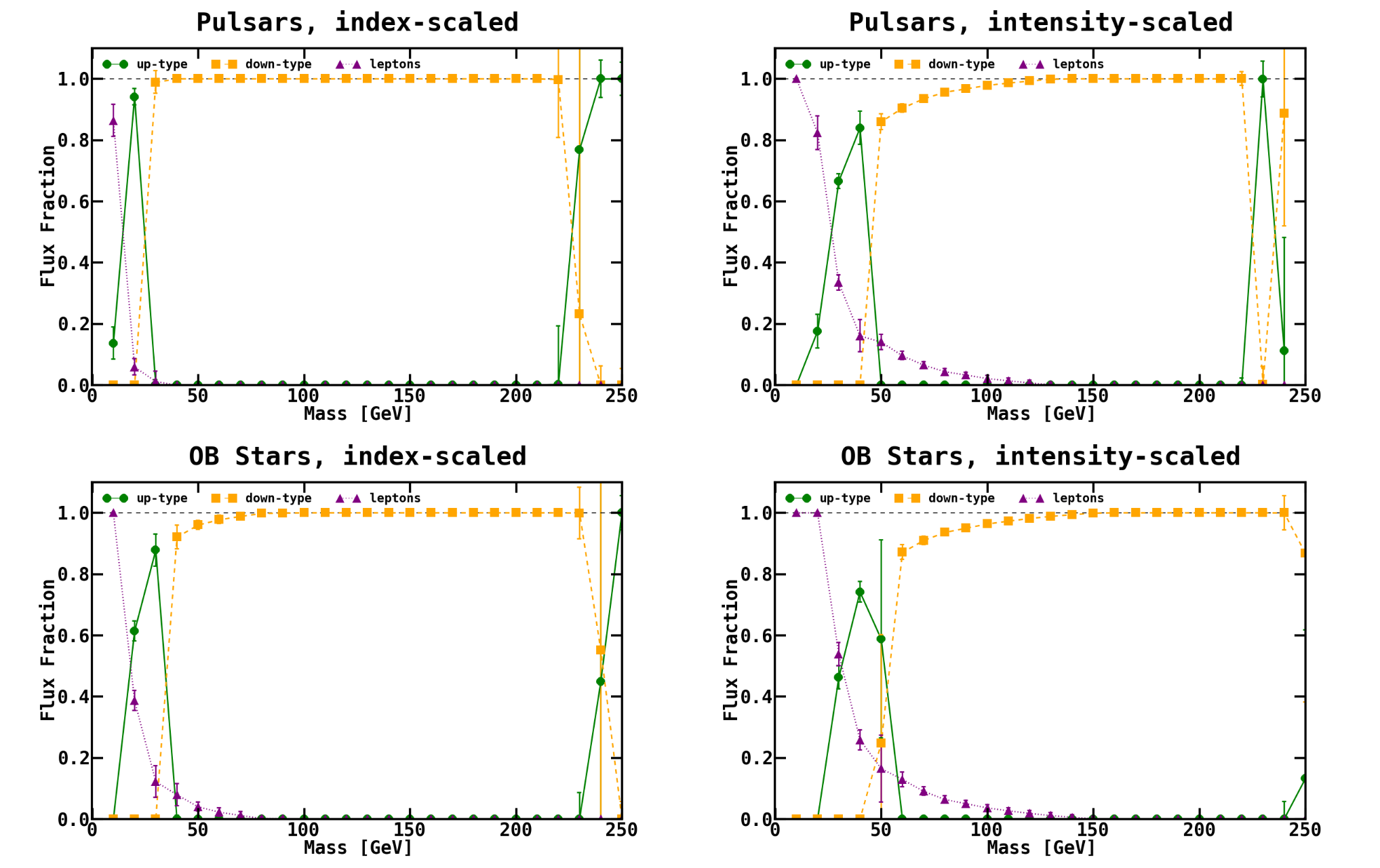}
\caption{Flux fraction for annihilation into up-type quarks, down-type quarks, and charged leptons, 
for the vector interaction model with NFW-c morphology. Results are shown for all four IEMs, as indicated.}
\label{fig:dmfracspin1}
\end{figure*}
%%%%%%%%%%%%%%%%%%%%%%%%%%%%%%%%%%%

%%%%%%%%%%%%%%%%%%%%%%%%%%%%%%%%%%%%%%%%%%%%%%
\begin{figure*}
\includegraphics[width=1.0\textwidth]{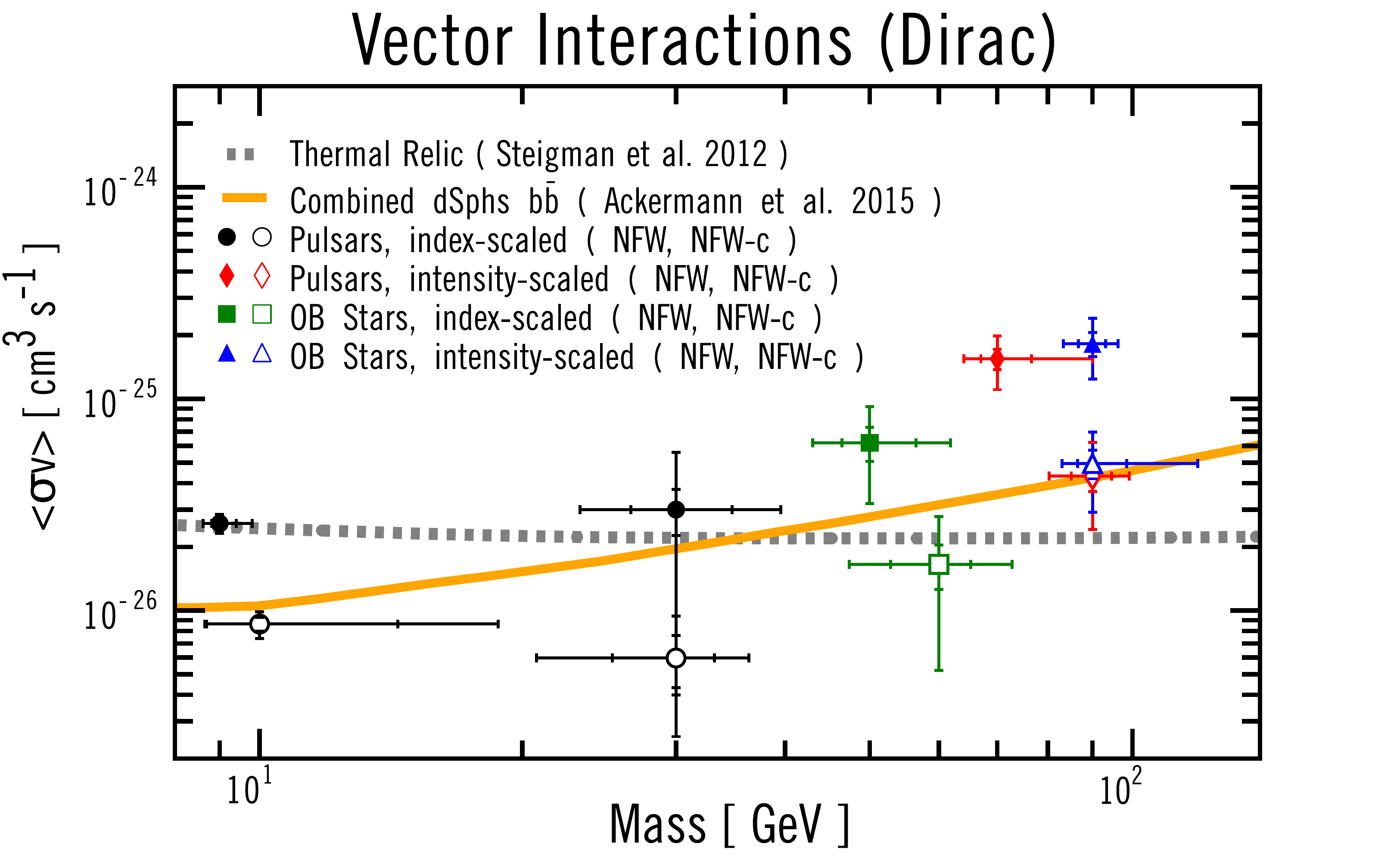}
\caption{Masses and cross sections for vector interaction models
(including one and two sigma uncertainties) for NFW and NFW-c DM profiles,
and the four IEMs, as indicated.  Also shown are the cross sections saturating the standard thermal relic density (grey dashed line)
and the \fermilat\ $95\%$ C.L. bounds from dwarf spheroidal galaxies, assuming $100\%$ annihilation into $b \overline{b}$.}
\label{fig:xsecspin1}
\end{figure*}
%%%%%%%%%%%%%%%%%%%%%%%%%%%%%%%%%%%%%%%%%%%%%%

The analysis for the vector-type DM interactions proceeds very similarly to the analysis of the pseudo-scalar interactions described above.
For each IEM and both NFW and NFW-c morphologies, the DM mass is scanned and the couplings to up-type quarks, down-type quarks, and
charged leptons is fit.  The results are presented in Figs.~\ref{fig:pslikespin1} and~\ref{fig:dmfracspin1}, respectively, for each IEM with the 
NFW-c profile (the results for the NFW profile are qualitatively similar.) 
Similarly to pseudo-scalar interactions, lower DM masses are favored by the index-scaled IEMs, compared to the intensity-scaled. 
However, in general, lower DM masses are favored for the vector interaction models than for the pseudo-scalar ones for the same IEM. 
In addition, because the coupling to SM fermions is assumed to be flavor-universal for the vector interaction model,
there is no sharp transition in behavior at the top quark mass.
For the {\em Pulsars, index-scaled} IEM, there are two close-to-degenerate minima in the likelihood profile, with the lower 
mass dominated by annihilations into leptons\footnote{For annihilations into 
leptons, secondary \gray\ emission via IC processes is neglected.  Note that for DM masses $\lesssim 10$~GeV,
IC photons are mainly produced at energies $< 1$~GeV \cite{Cirelli:2013mqa,Lacroix:2014eea}. }. 
The fitted values of $\langle \sigma v \rangle$ and the DM mass for each of the IEMs and DM profiles are shown in Fig.~\ref{fig:xsecspin1}. 

\section{Comparison with Other Searches}
\label{sec:compare}

As seen in sections~\ref{sec:spin0} and ~\ref{sec:spin1}, DM interpretations of the GC excess cover a broad range of 
masses ($\sim 10 - 200$ GeV) and $\langle \sigma v \rangle$, depending on the IEM, DM profile, and interaction type.  
One crucial avenue toward exploring a DM hypothesis for the excess is to compare the regions of parameter space best describing
the excess with the results from other searches for DM.  Null results of such searches can sharpen the target parameter space or
even exclude candidate explanations, whereas positive results could strengthen a DM interpretation of the excess and better
define the characteristics of candidate models.

\subsection{Indirect Searches}

For masses in the range $10 - 200$~GeV, the strongest constraints from indirect detection are generally from \fermilat ~observations of dwarf
spheroidal galaxies \cite{Ackermann:2015zua}.  These limits
appear to constrain the region relevant for explanations of the GC excess, but
are derived from less theoretically motivated
DM annihilation models where the DM annihilates into one species of SM fermion at a time.  As such, they do not precisely apply
to the models considered here,
although similar conclusions are likely.  The bound based on the assumption of $100\%$ annihilation into $b \overline{b}$,
corrected to account for Dirac (rather than Majorana) DM particles,
is shown on Figures~\ref{fig:xsec} and \ref{fig:xsecspin1} for reference.  The dwarf spheroidal bounds for annihilations into leptons  are not displayed in these figures. Although they would in principle be more pertinent to constrain our low mass, vector interaction results, they are still not adequate as the final state channel we consider here is an equal weight mixture of $e^+e^-$, $\mu^+\mu^-$,  $\tau^+\tau^-$ and therefore not directly comparable.

The limitations in the IEMs, modeling uncertainties in the dwarf halos \cite{Hayashi:2016kcy,Bernal:2016guq,Ichikawa:2016nbi,Klop:2016lug}, 
modifications to the particle physics model for DM \cite{Kopp:2016yji}, and
large uncertainties in the \jfac~for the GC \cite{Abazajian:2015raa}, all widen the relative uncertainties when 
confronting the parameters describing the
GC excess with the limits from observations of dwarf spheroidal galaxies.
Because of this, care must be taken when contrasting these limits with a DM interpretation of the GC excess.

The particle physics models under consideration also lead to annihilations producing anti-matter, such as positrons or anti-protons.  Positrons in particular
show excess production compared to naive expectations \cite{Adriani:2008zr,Aguilar:2013qda}, leading to
limits which do not significantly constrain the parameters for the GC excess \cite{Bergstrom:2013jra}.  
Recently, Ref.~\cite{Cuoco:2016eej} performed a detailed analysis of the anti-proton spectrum measured by AMS-02 \cite{Aguilar:2016kjl},
and also found an indication for an excess component roughly consistent with the parameter space describing a DM interpretation
of the GC excess (see also \cite{Giesen:2015ufa} for a less optimistic view).
The interpretation of CR anti-matter measurements is complicated by propagation, energy losses, and other modeling uncertainties related to particle fragmentation, as well as the spatial distribution of astrophysical sources. Consequently, the interpretation of these data in terms of DM is  unclear.

\subsection{Direct Searches}
 
Coupling to quarks implies coupling to hadrons, and thus is bounded from direct searches for DM scattering with heavy nuclei.  Models
with pseudo-scalar interactions map onto a scattering cross section which is both suppressed by the small velocities of DM in
the Galactic halo and are also spin-dependent.  As a result, the expectation is that the constraints from direct searches result in mild constraints.
In contrast, vector interactions lead to velocity-unsuppressed spin-independent scattering and are strongly constrained by direct searches.
For the vector models, which contribute to the spin-independent cross section $\sigma_{\rm SI}$, we follow the usual convention
mapping onto this quantity defined at zero relative velocity.
For pseudo-scalar interactions, 
we compute the integrated cross section for DM
scattering with a nucleon by
integrating over the recoil energy of the nucleus and the velocity of the DM, which we assume
follows a Maxwellian distribution,
using techniques
developed in \cite{Fitzpatrick:2012ix,Anand:2013yka,DelNobile:2013sia,Gresham:2014vja}, 
(specifically using the code presented in Ref.~\cite{DelNobile:2013sia}).  
This integrated cross section should be distinguished from usual spin-dependent cross section $\sigma_{\rm SD}$,
defined at zero velocity scattering, and is a more appropriate measure of scattering which is strongly velocity-dependent.

%%%%%%%%%%%%%%%%%%%%%%%%%%%%%%%%%%%%%%%%%%%%%%
\begin{figure*}
\includegraphics[width=1.0\textwidth]{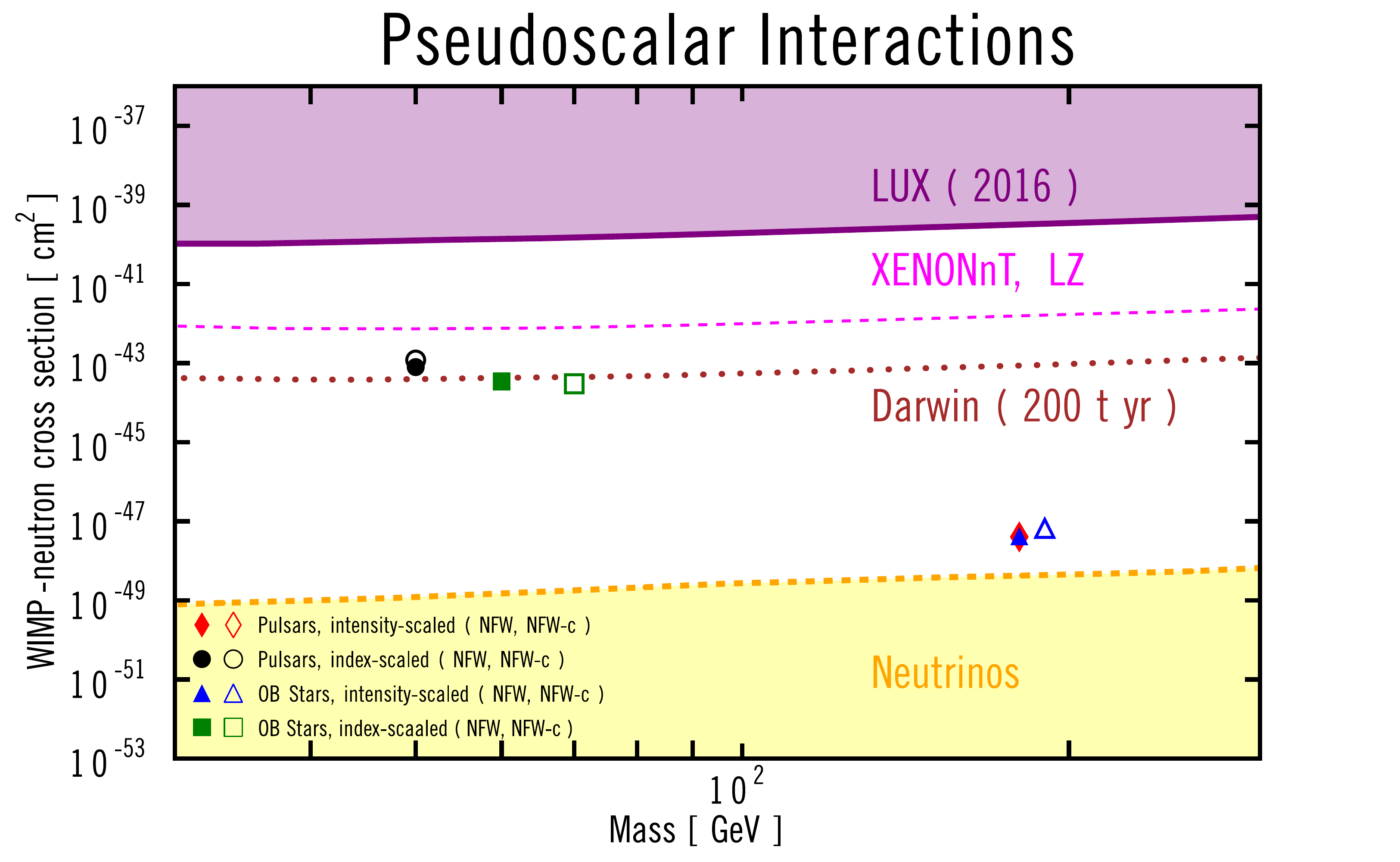}
\caption{ML points for the pseudo-scalar models, for each IEM and profile considered, as indicated, mapped into the plane of the DM mass and the integrated cross section,
as described in the text.  Also shown are current constraints from LUX (upper shaded region) 
and projections from XENONnT, LZ, and Darwin (dashed and dotted lines).  The lower shaded region indicates the neutrino floor.}
\label{fig:DD1}
\end{figure*}
%%%%%%%%%%%%%%%%%%%%%%%%%%%%%%%%%%%%%%%%%%%%%%

%%%%%%%%%%%%%%%%%%%%%%%%%%%%%%%%%%%%%%%%%%%%%%
\begin{figure*}
\includegraphics[width=1.0\textwidth]{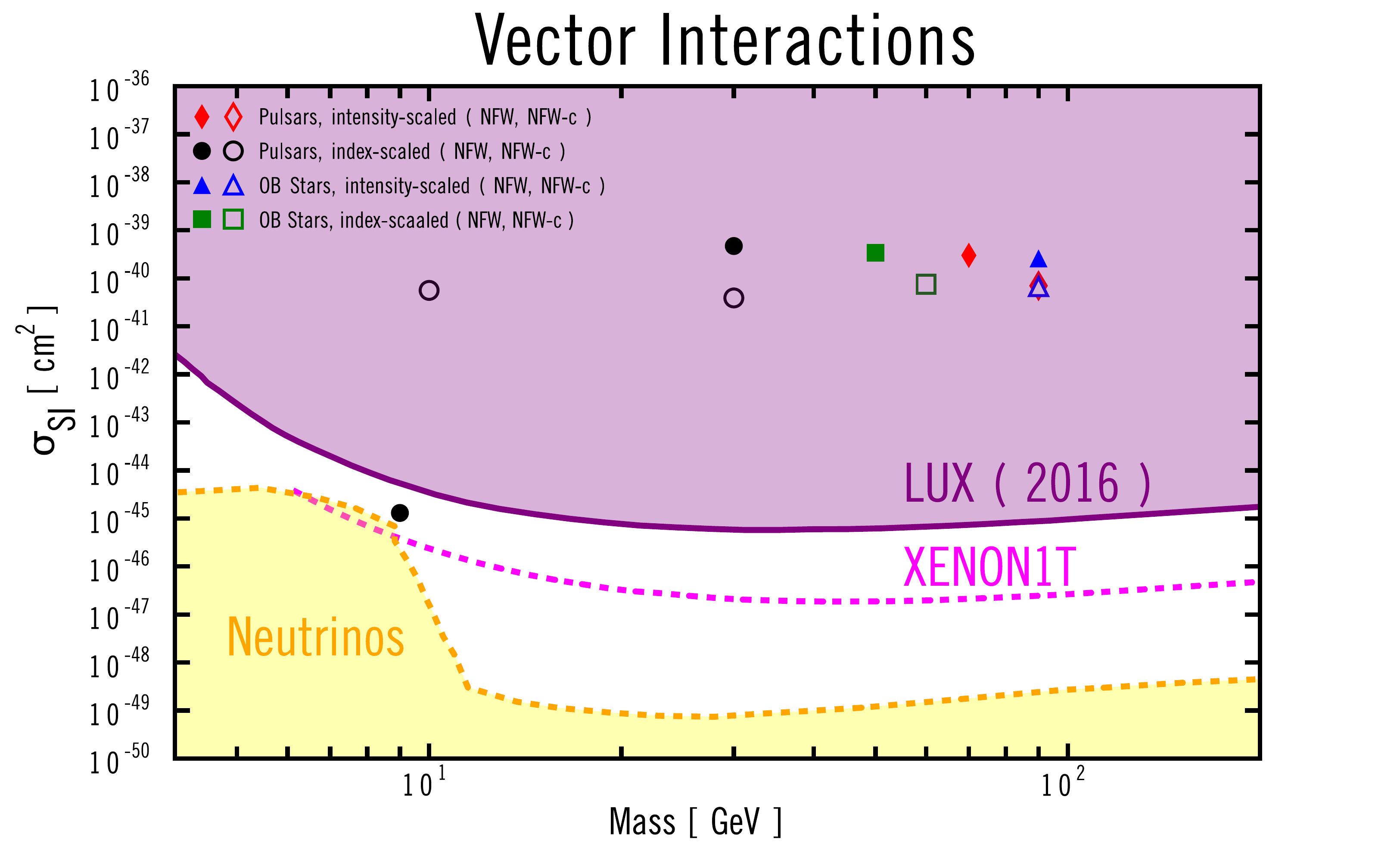}
\caption{ML points for the vector models, for each IEM and profile considered, as indicated, mapped into the plane of the DM mass and $\sigma_{\rm SI}$,
as described in the text.  Also shown are current constraints from LUX (upper shaded region) 
and projections from XENON1T (dashed line).  The lower shaded region indicates the neutrino floor.}
\label{fig:DD2}
\end{figure*}
%%%%%%%%%%%%%%%%%%%%%%%%%%%%%%%%%%%%%%%%%%%%%%

In Figs.~\ref{fig:DD1} and ~\ref{fig:DD2}, we show the ML points for the pseudo-scalar and vector models mapped into the WIMP-neutron 
spin-dependent integrated cross section, respectively, for each IEM and both NFW and NFW-c.  For
comparison, the limits from the LUX search for DM scattering with Xenon are presented \cite{Akerib:2016vxi}, 
also mapped into $\sigma_{\rm SI}$ or the integrated cross section for spin-dependent scattering with neutrons.
For the vector models, the limits from LUX easily exclude all of the ML points except for the point with dark
matter masses around $10$~GeV which annihilates predominantly into leptons for the {\em Pulsars, index-scaled}   IEM with NFW-c profile, which has
sufficiently small coupling to quarks that the scattering with nuclei is highly suppressed.
For the pseudo-scalar models, the predictions for the ML points lie well below below the LUX bounds, with the lower mass
points potentially probed long-term by Darwin \cite{Aalbers:2016jon}, while the higher mass points are slightly above the neutrino floor \cite{Billard:2013qya} and out of the reach of these experiments.
These results illustrate the importance of the IEM modeling and its influence on characterization of the putative signal, which can lead to
drastic differences in the expectations from complementary searches.
 
\subsection{Collider Searches}

Searches at the Large Hadron Collider (LHC) 
are more model-dependent, and can be classified based on the masses and couplings of the particles mediating the interaction.  When such particles
are heavy compared to the typical collider energies, they can be described by the same EFTs employed in this paper.
The results of searches in this regime are typically not competitive with direct searches except at masses far below those of interest to
describe the GC excess \cite{ATLAS:2012ky,Khachatryan:2014rra}.  For lighter mediating particles, the limits depend sensitively on
the specific couplings to the DM as well as to the SM fermions.  In particular, for values of the cross sections similar to what has been found in past characterizations of
the GeV excess, cases where a pseudo-scalar mediator's coupling to DM is significantly weaker than the coupling to quarks are mildly
constrained by LHC data, and the opposite limit is essentially unconstrained \cite{Fan:2015sza}.  Given the wide range of parameter space
(which is even larger for the specialized IEM analysis considered here), it seems possible that the LHC could eventually hope to observe
an excess consistent with a pseudo-scalar mediator interpretation if parameters are favorable. 
Similar remarks apply to the vector mediator models, although all but the {\em Pulsars, index-scaled} IEM with NFW-c profile are
already excluded by direct detection experiments.  This latter model is consistent with vanishing coupling to quarks, and thus is unlikely to
be excluded by searches at the LHC.

%%%%%%
\section{Summary}
\label{sec:conclusions}
%%%%%%

The excess of $\sim$~GeV \grays\ from the direction of the GC is an indication that there is something in the \gray\
sky beyond our current knowledge.  Whether this source ultimately proves to originate from DM annihilation or from a more conventional astrophysical
source still remains to be determined, and is likely to require further experimental input.  As part of this process, we have examined key
aspects of the putative signal using the specialized IEMs, developed by the \fermilat\ Collaboration~\cite{TheFermi-LAT}.
Our goal in characterizing potential DM explanations is to explore the implications from complementary searches, which can rule out or favor
a DM interpretation.

Our results illustrate the impact of interstellar emission modeling on the extracted characteristics of the excess and highlight the need for improved modeling
to capture a more realistic range of possibilities.  As far as the gross characteristics of the excess are concerned, we find 
 an offset of $\sim 0.5^\circ$ of the excess centroid from Sgr A* for all four IEMs considered.  We further find no significant evidence that the tail of the excess has a different spatial morphology than
the few GeV bump, with both high energy and low energy components favoring an NFW morphology compared to the other  morphologies we have considered.

We also consider flexible and realistic particle physics models for DM interacting with up-type quarks, down-type quarks, and charged
leptons, for two separate interaction types (pseudo-scalar and vector) leading to $s$-wave annihilation.  These theories are described by
EFTs, valid when the momentum transfer is small compared to the masses of the particles mediating the interactions -- to describe
annihilation, this implies the mediators are heavier than the DM itself.  We find that the choice of IEM has a large impact on the preferred
DM mass, annihilation cross section, and primary annihilation channel.  In particular, we identify regions with higher masses and
annihilation predominantly into top quarks.  Comparing the ML points in parameter space with direct and collider searches, we find that all of the
vector models aside from one at DM mass  $\sim 10$~GeV and annihilating into leptons are ruled out by null results from the LUX experiment.  The pseudo-scalar models
predict spin-dependent and velocity-dependent scattering with nuclei at a rate far below the current sensitivity, but in some cases within the grasp of
future planned experiments.  It would be interesting, but beyond the scope of this work, 
to extend our analysis beyond the EFT limit to the case of models where the DM can annihilate
directly into the mediator particles themselves.

The GeV excess is a compelling hint that there is more to learn about the Galaxy.  It is likely to take a combined effort of observation and interpretation to
unravel its nature.

\section*{Acknowledgements}

The authors are pleased to acknowledge conversations with D.~Finkbeiner, D.~Hooper, M.~Kaplinghat, T.~Slatyer, and C.~Weniger.
The work of CK and SM is supported in part by Department of Energy grant DE-SC0014431.
The work of TMPT and PT is supported in part by National Science Foundation grants PHY-1316792 and PHY-1620638.
GALPROP development is partially funded via NASA grants NNX09AC15G, NNX10AE78G, and NNX13AC47G.

\bibliography{FermiDMBIB}

\end{document}